\documentclass[11pt,letterpaper]{article}

\usepackage{amsmath}
\usepackage{mathtools}
\usepackage{amsfonts}
\usepackage{amssymb}
\usepackage{graphicx}
\usepackage{color}
\usepackage[left=3.5cm,top=3cm,right=3.5cm,bottom=3.5cm]{geometry}
\usepackage{multirow}
\usepackage{float}
\usepackage{array}
\usepackage{svg}
\usepackage{lineno}
\usepackage{setspace}
\usepackage[numbers]{natbib}
\usepackage{bm}
\usepackage{physics}
\usepackage{authblk}
\usepackage{caption}
\usepackage{subcaption}
\usepackage{hyperref}
\hypersetup{
colorlinks=true,
linkcolor=black,
filecolor=black,      
urlcolor=black,
citecolor=black,
}
\usepackage{mathrsfs}
\usepackage[version=4]{mhchem}

\usepackage[utf8]{inputenc}
\usepackage[T1]{fontenc}
\usepackage{lmodern}
\usepackage{mlmodern}

\title{Soft glassy materials with tunable extensibility}

\author[1]{Samya Sen}
\author[1]{Rubens R. Fernandes}
\author[1]{Randy H. Ewoldt\thanks{ewoldt@illinois.edu}}
\affil[1]{Department of Mechanical Science and Engineering\\University of Illinois Urbana-Champaign\\Urbana, IL 61801, USA}
\date{}

\begin{document}
	
\maketitle

\begin{abstract}
	
Extensibility is beyond the paradigm of classical soft glassy materials, and more broadly, yield-stress fluids. Recently, model yield-stress fluids with significant extensibility have been designed by adding polymeric phases to classically viscoplastic dispersions \cite{Nelson2018, AZN_CurrOpin2019, Bonn2022}. However, fundamental questions remain about the design of and coupling between the shear and extensional rheology of such systems. In this work, we propose a model material, a mixture of soft glassy microgels and solutions of high molecular weight linear polymers. We establish systematic criteria for the design and thorough rheological characterization of such systems, both in shear and in extension. Using our material, we show that it is possible to dramatically change the behavior in extension with minimal change in the shear yield stress and elastic modulus, thus enabling applications that exploit orthogonal modulation of shear and extensional material properties.
	
\end{abstract}

\section{\label{sec:intro}Introduction}
Yield stress fluids typically exhibit a reversible transition between a solid-like and a fluid-like state around a critical applied stress. This behavior can originate from several microstructural interactions \cite{Piau_cpol2007, CoussotReview2014, Bonn_YS_JNNFM2016, BonnManneville2017}, such as the repulsive potential between particles in soft glasses (emulsions and microgel dispersions). In these materials, particles are dispersed in a liquid continuous phase at relatively high volume fractions, preventing thermal relaxation of the microstructure, leading to glassy or jammed states \cite{sollich1998rheological}. This leads to solid-like elasticity from soft particle contact forces and soft glassy materials are therefore often studied within the paradigm of yield-stress (elastoviscoplastic) fluids.

Designing extensibility into yield-stress fluids is of particular interest in applications such as rapid prototyping and direct-ink writing, where a combination of yield stress and extensibility is beneficial for optimized printing performance \cite{Rauzan2018, AZN_CurrOpin2019, ChaiRHE_ARFM}. Applications such as droplet and jet impacts on surfaces, e.g.\ as motivated by fire suppression spray coatings, may also utilize fluids with tunable shear and extensional viscosities \cite{SSRHE_DFD2019, ChaiRHE_ARFM, keshavarz2012elastic, hsu2011role, Sen_PhDThesis}. 



Despite the apparent importance of extensible yield-stress fluids, very few studies have reported on the formulation and testing of model yield-stress fluids with significant and tunable extensibility. Only recently, materials based on emulsions have been designed with significant extensional properties \cite{AZN_SM2017,Nelson2018,Rauzan2018,AZN_CurrOpin2019}. However, these materials suffer from experimental artifacts that hinder their characterization in both shear and extension, e.g.\ edge fracture in shear rheology. Also, such materials can be difficult to formulate with consistent properties such as shear yield stress or elastic modulus, owing to the imprecise control over the size distribution of the dispersed phase droplets. Additionally, accurate characterization of the yield stress in shear and extensional flows has been a topic of debate in the recent literature. Although soft glassy materials such as Carbopol and emulsions are often assumed to follow the von Mises yield criterion  \cite{Carbopol_Ovarlez, shaukat2012shear}, recent observations indicate that this may not hold for yield stress fluids in general \cite{sica2020mises}.

In this work, we design, formulate, and characterize a model extensible yield-stress fluid that overcomes the limitations described above. We use a Carbopol microgel suspension, a soft glassy material that is widely used in the literature as a model yield-stress fluid to study microstructural, rheological, and flow physics \cite{Piau_cpol2007,CoussotReview2014,Bonn_YS_JNNFM2016,BonnManneville2017,Kamani_PRL2021,LuuForterre2009,LuuForterrePRL2013,BCB_PhysFluids2015,SSRHE_JFM2020}, and impart extensibility to it by making physical mixtures with solutions of a high-molecular weight poly(ethyleneoxide) (PEO) linear chain polymer. We perform a thorough rheological characterization in both shear and extension to study the variation in extensional and shear rheological properties. We show that key properties in shear, such as yield stress and linear viscoelastic storage modulus, do not change by more than 10-20\% upon addition of PEO. On the other hand, filament stretching extensional tests show that addition of polymers changes the extensional properties, such as filament failure strain, by 200-300\%. These results strongly support the usefulness of this system as a model yield-stress fluid with tunable extensibility.


\section{\label{sec:matmeth}Materials and methods}

\subsection{\label{subsec:materials}Materials used}
\begin{figure}[!ht]
\centering
\includegraphics[width=0.75\linewidth]{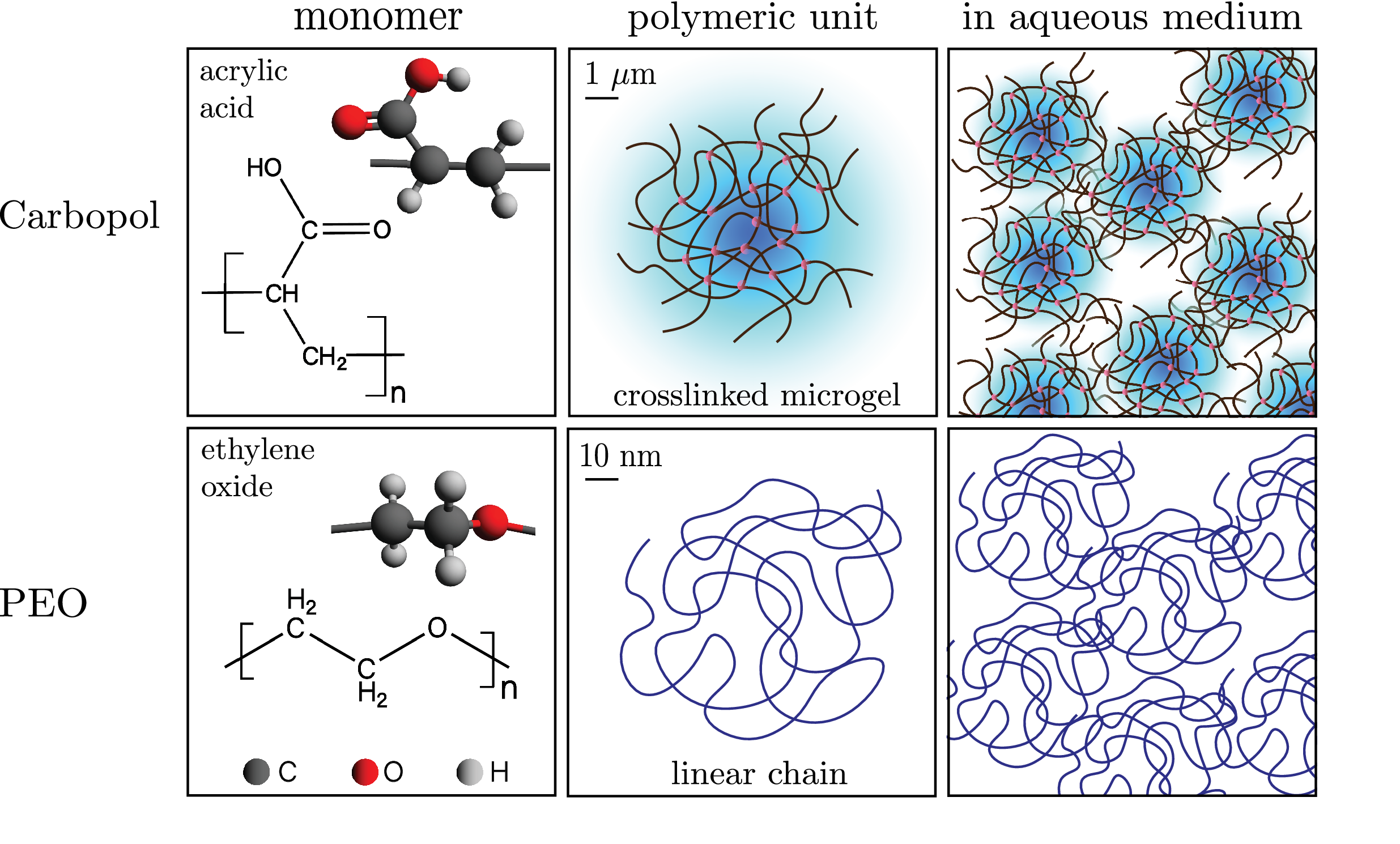}
\caption{\label{fig:cpolpeo-struct} Hierarchical structures at different length scales for Carbopol and PEO in water. The orders of magnitude separation in length scales is indicated.}
\end{figure}

Carbopol is widely employed to study and model the rheological properties and flow physics of yield-stress fluids. It is a soft, glassy (repulsion dominated) fluid \cite{Carbopol_Cates,Carbopol_Kim,Piau_cpol2007,Carbopol_Ovarlez,Nelson2018}. Carbopol particles are microgels made of cross-linked poly(acrylic acid) chains. When dispersed in water and the $p$H raised to 6.5-7.5 using a base, the microgel particles swell, leading to a crowded microstructure, forming a transparent soft glassy material \cite{Piau_cpol2007,Carbopol_Ovarlez}. Carbopol powder (Carbopol 980, Lubrizol) was thoroughly mixed in de-ionized water for 6~h using a magnetic stirrer at 200~r.p.m. After mixing, 1~M aqueous \ce{NaOH} solution was added dropwise until the $p{\rm H}$ was stable around $7 \pm 0.10$. The mixture was stirred gently at 50~r.p.m.\ for 12~h to homogeneously disperse the base within the sample volume. The samples were tested either in their pure form, or in a mixture with PEO solution. A schematic of the structure of Carbopol at various length scales is shown in Fig.~\ref{fig:cpolpeo-struct}.

PEO consists of ethylene oxide monomers which are polymerized into linear chains. The samples were made from PEO that comes in a fine powder form (Sigma Aldrich, $M_{\rm v} \simeq$~4,000~kDa), dissolved in a 19:1 de-ionized water-ethanol co-solvent. Requisite amounts of the powder were dissolved into the co-solvent by vigorously mixing in a beaker with a magnetic stir rod and stir plate at 200~r.p.m.\ for 12~h and a homogeneous viscoelastic solution was obtained.

Polymer solutions are in the dilute regime when chains do not overlap or interact significantly. The semi-dilute regime is when overlap between hypothetical spheres enveloping the chains start interacting directly \cite{Bailey1959,Bailey_book}. The overlap concentration, $c_{\rm PEO}^*$, is used to approximate the onset of this regime, and is defined as $c_{\rm PEO}^*=1/[\eta]_0$, where $[\eta]_0$ is its intrinsic viscosity \cite{Bailey1959,Bailey_book}. For PEO, it is empirically related to the average molecular weight $M_\text{v}$ of the polymer as $[\eta]_0 \simeq 0.0125 M_\text{v}^{0.78}$ \cite{Bailey1959,Bailey_book}. Using this, the overlap concentration for the 4,000~kDa PEO used in this work was calculated as $c_{\rm PEO}^* \approx 0.0567~\text{wt}\%$. Among the samples used in our study, only the 0.01~wt\% sample is in the dilute regime, whereas the 0.10, 0.50 and 1.00~wt\% samples are in the semi-dilute to concentrated regime. As the PEO concentration increases, we will see in the following sections that the extensibility of Carbopol+PEO also increases dramatically, while the steady and viscoelastic shear properties do not change as much. A schematic of PEO structure at various length scales in shown in Fig.~\ref{fig:cpolpeo-struct}.

The swelled Carbopol microgel units form clusters that lead to a glassy state due to their repulsive interactions, with the diameter of each swollen particle $\mathcal{O}(1)~\mu{\rm m}$ \cite{Piau_cpol2007} (i.e.\ the polymeric unit in Fig.~\ref{fig:cpolpeo-struct}). PEO molecules, on the other hand, are orders of magnitude smaller than the microgel particles, with $R_{\rm g} \sim \mathcal{O}(10)$~nm for a single chain polymeric unit. Consequently, we hypothesize that the PEO molecules may not influence the physical interactions between the microgel particles significantly, although they are expected to alter the behavior of the mixture in other ways, which will be discussed later.

\begin{figure}[!ht]
\centering
\includegraphics[width=0.75\linewidth]{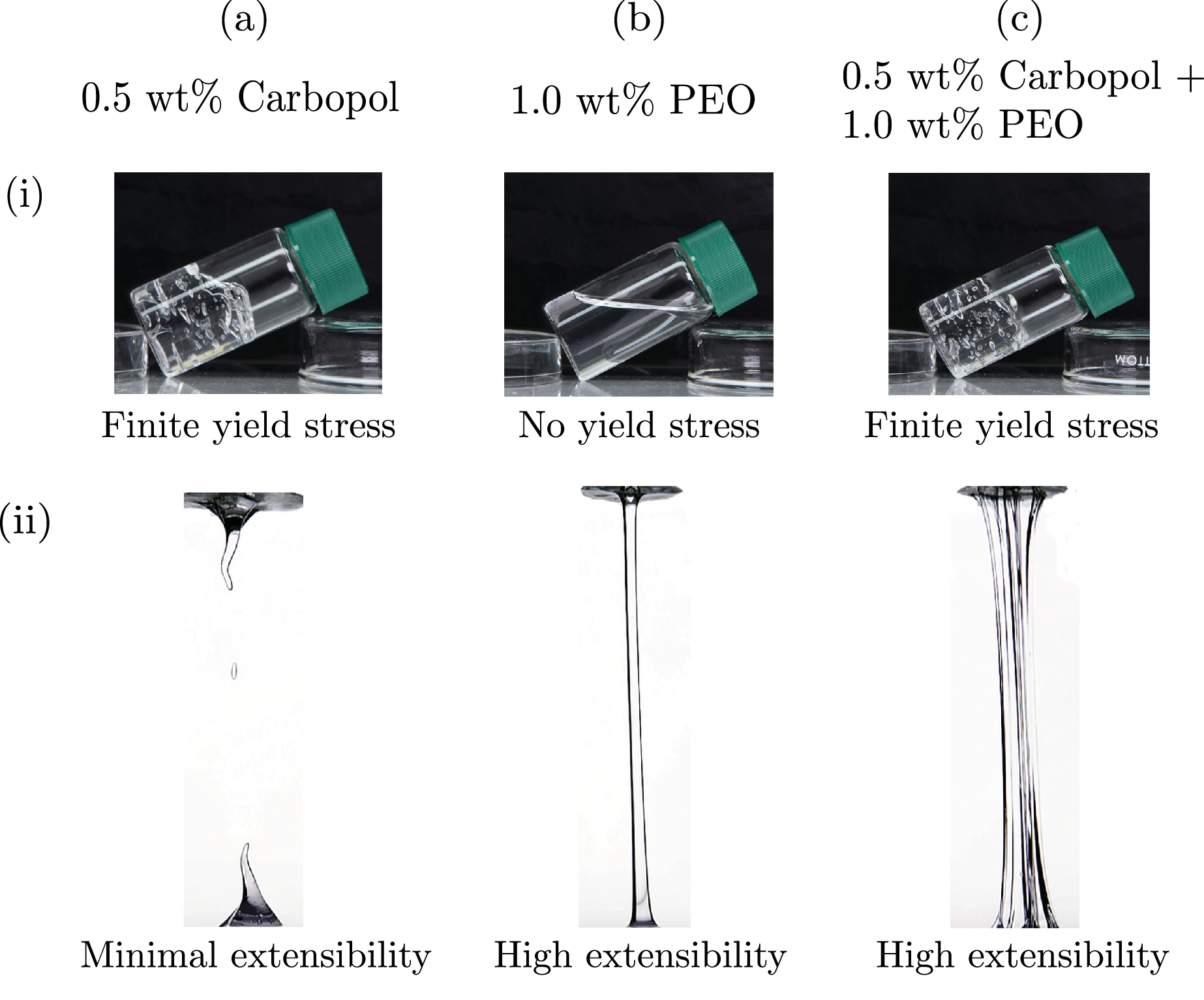}
\caption{\label{fig:matmeth-tilt_fiser}Plausible visual evidence for (i) the presence of a yield-stress for Carbopol versus an absence in PEO (over the timescale of observation in the images, $\sim 1$~h), and (ii) the insignificant extensibility of Carbopol versus prominent extensional behavior of PEO; samples stretched between plates under strain rates of $\dot{\epsilon} \sim \mathcal{O}(10)~{\rm s}^{-1}$. Upon mixing, the fluid retains the properties of both its constituent components, having both a yield stress and large extensibility.}
\end{figure}

The extensible yield-stress fluid (ExYSF) was obtained by mixing the fully prepared Carbopol and PEO formulations. Depending on the target weight fractions of Carbopol and PEO in the final ExYSF sample, requisite amounts of each material were mixed in a stirrer at 50~r.p.m.\ for 6~h to make the mixture as homogeneous as possible. Containers were sealed during stirring to minimize solvent evaporation, and very high shear rates were avoided during mixing to prevent scission of the chains in either material.

As can be seen from the protorheology~\cite{Hossain2023} photos in Fig.~\ref{fig:matmeth-tilt_fiser} (a-i), Carbopol samples are transparent and the yield stress is evident from its ability to support its own weight when the container is tilted \cite{TanRHE}. However, pure Carbopol dispersions are not appreciably extensible, Fig.~\ref{fig:matmeth-tilt_fiser} (a-ii). On the other hand, PEO solutions in water are clear, and flow at infinitesimal stresses, showing no yield stress as is evidenced by it taking the shape of the container (b-i). The samples show significant stretching between two parallel plates, and thus have very high extensibility (b-ii). When the Carbopol suspension is mixed with the PEO solution, the resultant material shows interesting features: it can support its own weight (c-i), indicating a yield stress behavior, and also shows high extensibility (c-ii). In the following sections, we support these qualitative observations with quantitative rheological measures of characterizing these features in shear and extension, focusing on the effect of added PEO.

\subsection{\label{subsec:methods}Rheometry methods}
Shear rheometry was performed on an ARES-G2 strain-controlled, separated motor-transducer rheometer (TA Instruments), using a 25~mm parallel-plate geometry, with adhesive-backed 600~grit sandpaper to mitigate wall slip. All tests were conducted at $25^\circ{\rm C}$ using an Advanced Peltier System to regulate the temperature. This was chosen to match the ambient temperature of the room, since controlling temperature in extension experiments is not possible owing to the large aspect ratio (and hence large exposed surface area) of the samples in filament stretching.

We use filament stretching extensional rheometry (FiSER) to obtain the extensional properties of our samples \cite{YaoMcKinley_JNNFM1998,McKinleySridhar_AnnuRev2002,Nelson2018, koeppel2018extensional}. In these tests, a cylindrical liquid bridge is initially formed between two circular end-plates. The plates are then moved apart in a prescribed manner such that the fluid sample is subjected to a predominantly extensional deformation \cite{McKinleyHassager_JoR1999,BachHassager_JNNFM2002,McKinleySridhar_AnnuRev2002,BachHassager_JoR2003}. With time, the accumulated strain in the sample increases; the filament starts to neck as it is increasingly stretched, then dramatically thins at its cross-section where necking first occurred. This is finally followed by pinch-off failure. The pinch-off or filament breakup is primarily driven by the boundary conditions at the end plates, although capillary-driven thinning does play a minor role in our test conditions. In our tests, we used 8~mm parallel plates (radius $R_0=4$~mm) to stretch fluid samples on the ARES-G2 rheometer in ambient conditions, with an initial gap of $L_0=4$~mm. This gives an initial sample aspect ratio of $\Lambda_0 \equiv L_0/R_0 = 1$, consistent with previous works on extensible yield-stress fluids \cite{Nelson2018,Rauzan2018}. This aspect ratio was chosen to reduce radial flow effects \cite{YaoMcKinley_JNNFM1998}. The upper plate of the rheometer was moved up to increase the gap $L(t)$ between the plates and stretch the sample at a constant nominal true strain rate (Hencky strain rate) of $\dot{\epsilon}_0 \equiv \dot{L}(t)/L(t) =0.2~{\rm s}^{-1}$. This requires that the upper plate move at an exponentially increasing linear velocity, which places a limit on the maximum Hencky strain rate that can be applied to the sample \cite{Nelson2018}. After loading the sample, the residual normal stress was allowed to relax to the noise level of the instrument force transducer ($F_{\rm min} \approx 1~{\rm mN}$), and the axial force on the upper plate was measured during the experiment for further calculations.


\section{\label{sec:shear}Effect of PEO on steady shear properties}

\begin{figure}[!ht]
\begin{minipage}[!ht]{0.49\textwidth}
	\centering
	(a)
	\includegraphics[width=\linewidth]{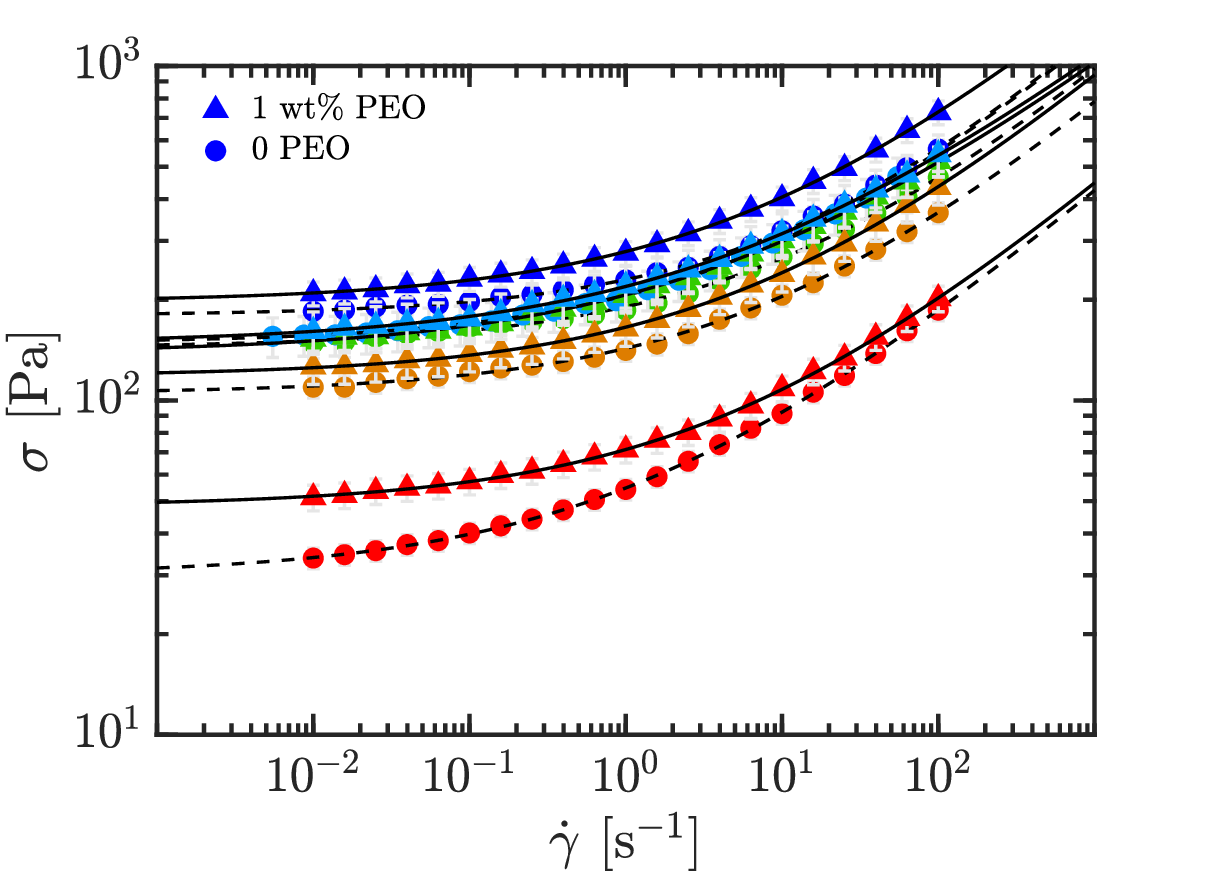}
\end{minipage}
\begin{minipage}[!ht]{0.49\textwidth}
	\centering
	(b)
	\includegraphics[width=\linewidth]{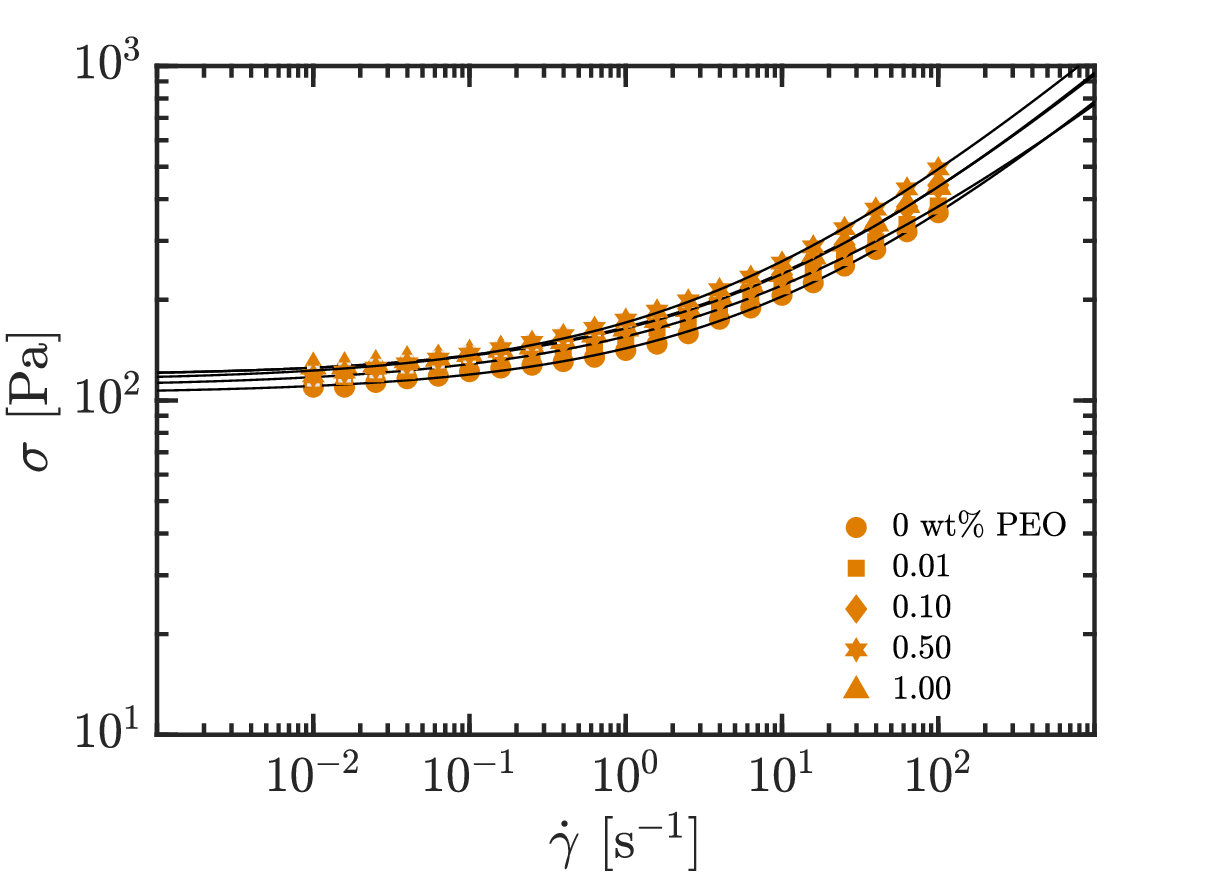}
\end{minipage}
\caption{\label{fig:flow-flowcurves}Steady shear data for various Carbopol+PEO samples. The fit lines are for the Herschel-Bulkley model, Eq.~(\ref{eq:steadyshear-eq1}). (a) All wt\% Carbopol with 0 or 1~wt\% PEO added. (b) 0.50~wt\% Carbopol with 0, 0.01, 0.10, 0.50, and 1.00 wt\% PEO added. The triangles refer to the Carbopol samples with 1~wt\%, and circles to those with 0 PEO. Each color corresponds to a different wt\% of Carbopol (progressively red to blue colors for 0.25, 0.50, 1.00, 1.50, 2.00~wt\% respectively).}

\end{figure}

The steady shear data for all Carbopol+PEO samples are shown in Fig.~\ref{fig:flow-flowcurves}.

From Fig.~\ref{fig:flow-flowcurves}(a), both types of formulations, with and without PEO, have an apparent yield stress, evident from the plateauing stress at low shear rates. Secondly, the shear stress is an increasing function of shear rate, with the material stress gradually transitioning from the plateau yielding to a flowing region as the rate increases. The same yielding and flow behavior is also observed in creep tests, data for which is shown in the Supplementary Information (SI). These signatures are typical of yield-stress fluids, and adding PEO to Carbopol does not alter these characteristic features. The behavior at low shear rates is not very different for a given Carbopol concentration at 0 \emph{versus} 1~wt\% PEO, except for the lowest Carbopol loading of 0.25~wt\%. The high shear rate behavior is also very similar. The effect of added PEO, while finite, is not large.

\begin{figure}[!ht]
\centering
\includegraphics[width=0.5\linewidth]{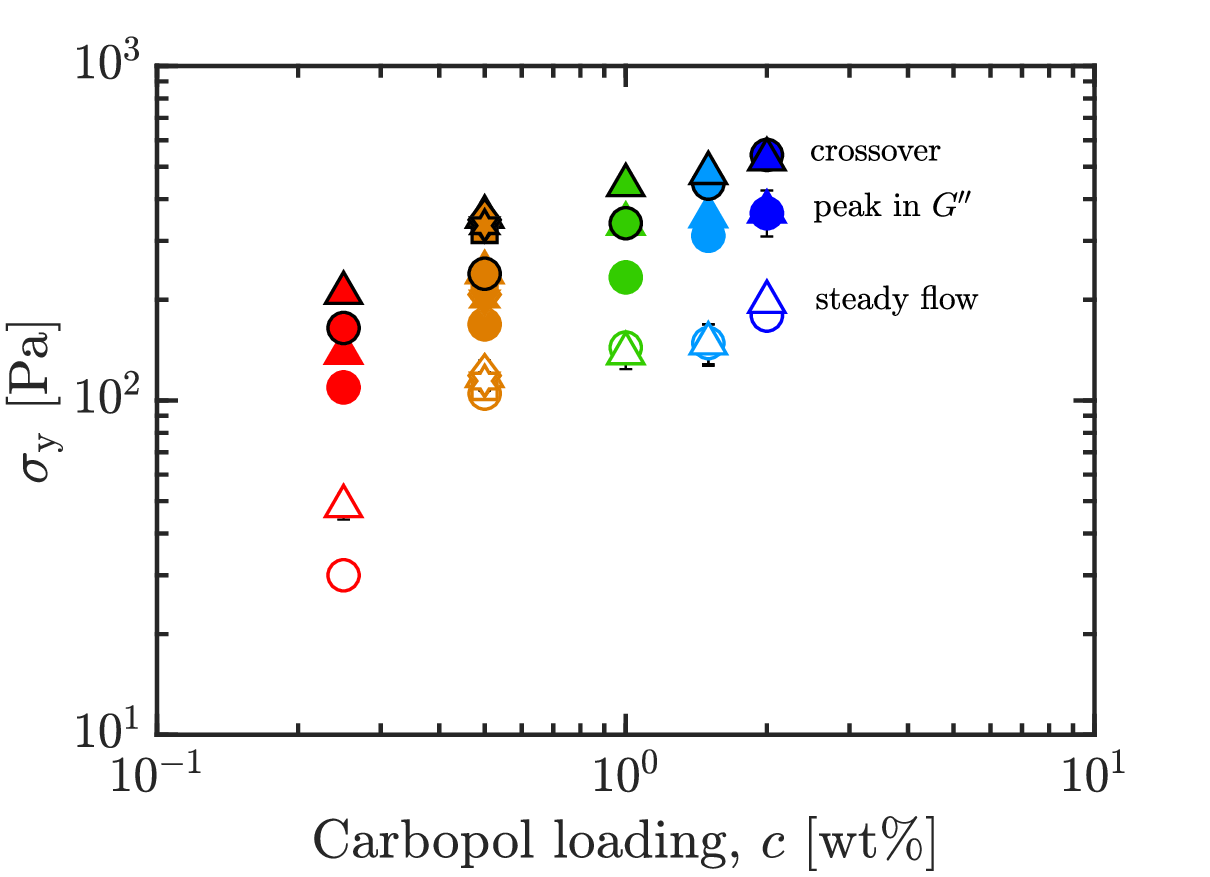}
\caption{\label{fig:ys-conc}Shear yield stress $\sigma_{\rm y}$ obtained from steady flow data and LAOS metrics ($G^\prime,G^{\prime\prime}$ crossover, and peak in $G^{\prime\prime}$), \emph{versus} Carbopol loading $c$, for 0 (circles) and 1~wt\% (triangles) PEO, all wt\% Carbopol. Co-plotted are results for 0.50~wt\% Carbopol with 0, 0.01, 0.10, 0.50, and 1.00 wt\% PEO added (polygonal symbols). Only the steady shear yield stress is used hereafter. (See Fig.~\ref{fig:LAOS-Gvstrain} for crossover and peak in $G^{\prime\prime}$ primary data from LAOS.)}
\end{figure}

This is seen more clearly from Fig.~\ref{fig:flow-flowcurves}(b), where the yield stress and high shear scaling do not change appreciably for various PEO wt\% added to 0.50~wt\% Carbopol. Further insight can be gained by fitting this data to the Herschel-Bulkley (HB) model \cite{DPL_vol1,Macosko:1994,CoussotReview2014,BonnManneville2017}, given by
\begin{align}\label{eq:steadyshear-eq1}
\sigma = \sigma_{\rm y} + K\dot{\gamma}^n \equiv \sigma_{\rm y}\left[ 1 + \left( \frac{\dot{\gamma}}{\dot{\gamma}_{\rm crit}} \right)^n \right],
\end{align}
where $\sigma_{\rm y}$ is the dynamic steady shear yield stress, $K$ is the consistency index, and $n$ is the flow index. An equivalent form of the model, introduced by Dinkgreve \emph{et al.} \cite{DinkgreveBonn2015}, is also shown, which involves a critical shear rate, $\dot{\gamma}_{\rm crit}$, instead of $K$ as a parameter \cite{DinkgreveBonn2015,AZN_SM2017,Nelson2018,Caggioni2020}, to avoid the problem that the dimensions of $K$ depend on $n$. Steady shear data fits to the HB model for the ExYSF are shown in Fig.~\ref{fig:flow-flowcurves} as solid lines (Carbopol with 1~wt\% PEO) and dashed lines (pure Carbopol). The dynamic yield stress is shown for all samples in Fig.~\ref{fig:ys-conc} as a function of Carbopol loading.

Consistent with our hypothesis, while $\sigma_{\rm y}$ increases with Carbopol concentration $c$, the addition of PEO does not lead to large variations in $\sigma_{\rm y}$ for $c>0.50$~wt\%. The yield stress of the sample with the smallest Carbopol loading ($0.25$~wt\%, red symbols in Fig.~\ref{fig:flow-flowcurves}) was more affected by the added PEO. This likely results from the comparatively smaller yield stress displayed by this sample, meaning that crowding effects are not dominant over the effects of added PEO. This is true for all estimates of $\sigma_{\rm y}$ obtained in our experiments, whether from steady shear or oscillatory shear flows. To obtain yield stress from oscillatory rheology, the stress at the crossover of $G^\prime,G^{\prime\prime}$ or the stress at the peak in $G^{\prime\prime}$ in LAOStrain may be used \cite{Bonn_YS_JNNFM2016,DonleyRogers_JNNFM2019}, among other possibilities. In Fig.~\ref{fig:ys-conc}, we co-plot $\sigma_{\rm y}$ obtained from various metrics using data from LAOS experiments \cite{AsheshGaurav_SM2019,DonleyRogers_JNNFM2019}, and the effect of added PEO is not very significant across any measure of $\sigma_{\rm y}$, and they all follow the same trend with Carbopol loading. Details of oscillatory shear data are given in Sec.~\ref{sec:shear-osc}.

	At low concentrations, addition of PEO to the Carbopol system may change the effective volume fraction of particles, enhancing interparticle interactions between microgel particles, thus increasing the yield stress compared to the pure microgel system. But this effect is overshadowed by the microgel particles themselves getting heavily jammed and producing a much larger yield stress at higher concentrations. This explains the reason for variation between 0 and 1~wt\% PEO for low Carbopol loading, yet a consistent yield stress across this range of PEO concentration at higher Carbopol loadings.
	
	We have also found that the first normal stress difference in shear, $N_1$, is measurable in our Carbopol samples during steady shear tests. The data is shown in the SI. We observe that $N_1$ increases with Carbopol loading for a given rate, and for each Carbopol loading, $N_1$ increases with the addition of 1~wt\% PEO. Upon PEO addition, $N_1$ increased by a factor of 1.5-2 in the range of shear rates where the data was outside the noisy regime. Interestingly, this increase is also comparable to the increase in extensibility (2-3 times increase in failure strains) for Carbopol+PEO when compared with pure Carbopol, as will be shown in the next section. This increase in $N_1$ suggests that it is correlated to the increased extensibility.
	
	To summarize, we find that the steady shear yield stress is only marginally changed upon adding PEO, except at the lowest Carbopol loading, while increasing shear normal stress differences significantly. In the next section, we will study the effect of PEO on the extensibility of Carbopol systems.
	


	\section{\label{sec:extension}Effect of PEO on extensional properties}
	In FiSER tests, the axial force $F_z$ on the end plates can be measured during the filament stretching process, from which tensile stresses and apparent extensional viscosities can be calculated \cite{McKinleyHassager_JoR1999,McKinleySridhar_AnnuRev2002,BachHassager_JNNFM2002,BachHassager_JoR2003,Valette2019,Tsamopoulos_EVP2020}. While material functions are difficult to measure precisely for FiSER tests with these materials, owing to the necking, non-ideal deformation history, and forces near the measurement limit, there are measurable quantities that are useful for practical applications. One such quantity is the extensional strain-to-break \cite{Nelson2018,Rauzan2018,Sen_PhDThesis}, as also used in solid mechanics \cite{Callister}, which is the engineering strain at which the filament breaks. It is calculated as
	\begin{align}\label{eq:eSTB-definition}
		\varepsilon_{\rm b} \equiv \frac{\Delta L_{\rm b}}{L_0} = \frac{L_{\rm b} - L_0}{L_0},
	\end{align}
	where $L_{\rm b}$ is the filament breaking length, and $L_0$ is the initial filament length. Similar criteria connected to the filament breaking length $L_{\rm b}$ have also been used in the literature \cite{James_RheolActa2006}.
	
	\begin{figure}[!ht]
		\includegraphics[width=\linewidth]{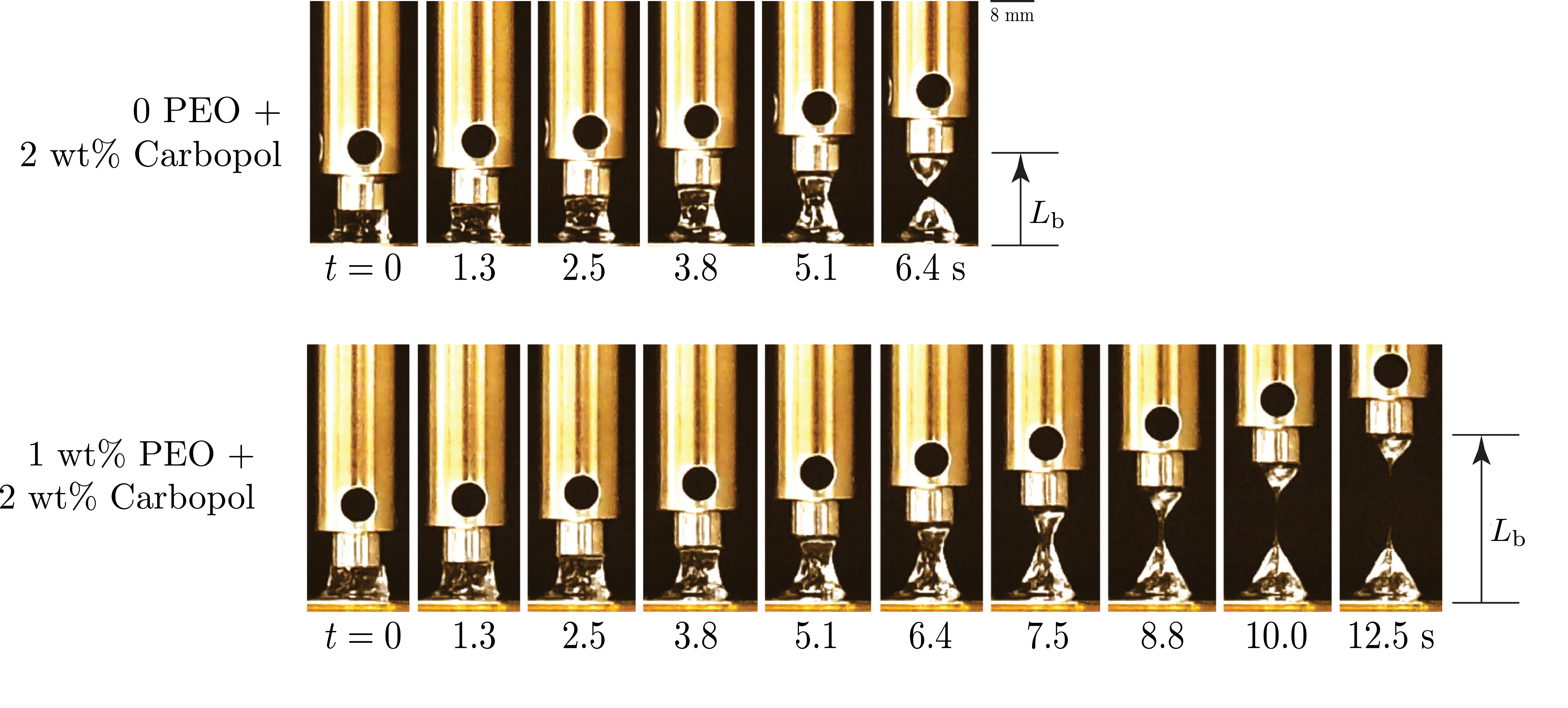}
		\caption{\label{fig:fiser-ares}Filament stretching tests and strain-to-break observations for 2~wt\% Carbopol with 0 and 1~wt\% PEO. These tests were done at a fixed applied nominal Hencky strain rate, $\dot{\epsilon}_0=0.2~\text{s}^{-1}$. The samples with added PEO display larger $L_{\rm b}$ (and thus larger $\varepsilon_{\rm b}$), as seen from the delayed filament pinch-off.}
	\end{figure}
	
	
	Filament breakup criteria (i.e., the stresses and strains at filament failure) are often used to quantify data obtained from extensional tests conducted with viscoelastic materials. Either necking or failure strain can thus be used as a metric for gauging how extensible the material is \cite{Fielding_PRL2011,HoyleFielding_JoR2016_1,HoyleFielding_JNNFM2017}. The engineering strain-to-break $\varepsilon_{\rm b}$ is a function of both time $t$ and the nominal strain rate $\dot{\epsilon}_0$ applied to the plates. The failure of highly elastic solids such as metals is governed by the accumulated strain, meaning that in these cases $\varepsilon_{\rm b}$ depends weakly on the nominal strain rate imposed on the material. Elastoviscoplastic fluids such as the ExYSF used in this work, on the other hand, show both elastic (strain dependent) and viscous (rate dependent) dynamics, so the extensional strain-to-break is expected to depend not only on the accumulated strain, but also on the rate at which the strain is accumulated.
	
	Fig.~\ref{fig:fiser-ares} shows the results of filament stretching tests for 2~wt\% Carbopol with 0 and 1~wt\% PEO added. From the figure, we see that the filament breaking length $L_{\rm b}$ is greater for the sample with 1~wt\% PEO ($L_{\rm b} \approx 7.30L_0$) when compared to the pure Carbopol sample ($L_{\rm b} \approx 2.95L_0$). These correspond to a filament breaking engineering strain of $\varepsilon_{\rm b} \approx 630\%$ for Carbopol with 1~wt\% PEO \emph{versus} $\varepsilon_{\rm b} \approx 195\%$ for pure Carbopol. From these images, the dramatic differences in extensional properties of the two fluids as a result of the added polymer becomes very clear. Since the same nominal extension rate was used for both experiments, a larger extensional strain-to-break also corresponds to a delayed failure time $t_{\rm b}$ of the material $(t_{\rm b} \sim \ln(1+\varepsilon_{\rm b})/\dot{\epsilon}_0)$.
	
	\begin{figure}[!ht]
		\centering
		\includegraphics[width=0.5\linewidth]{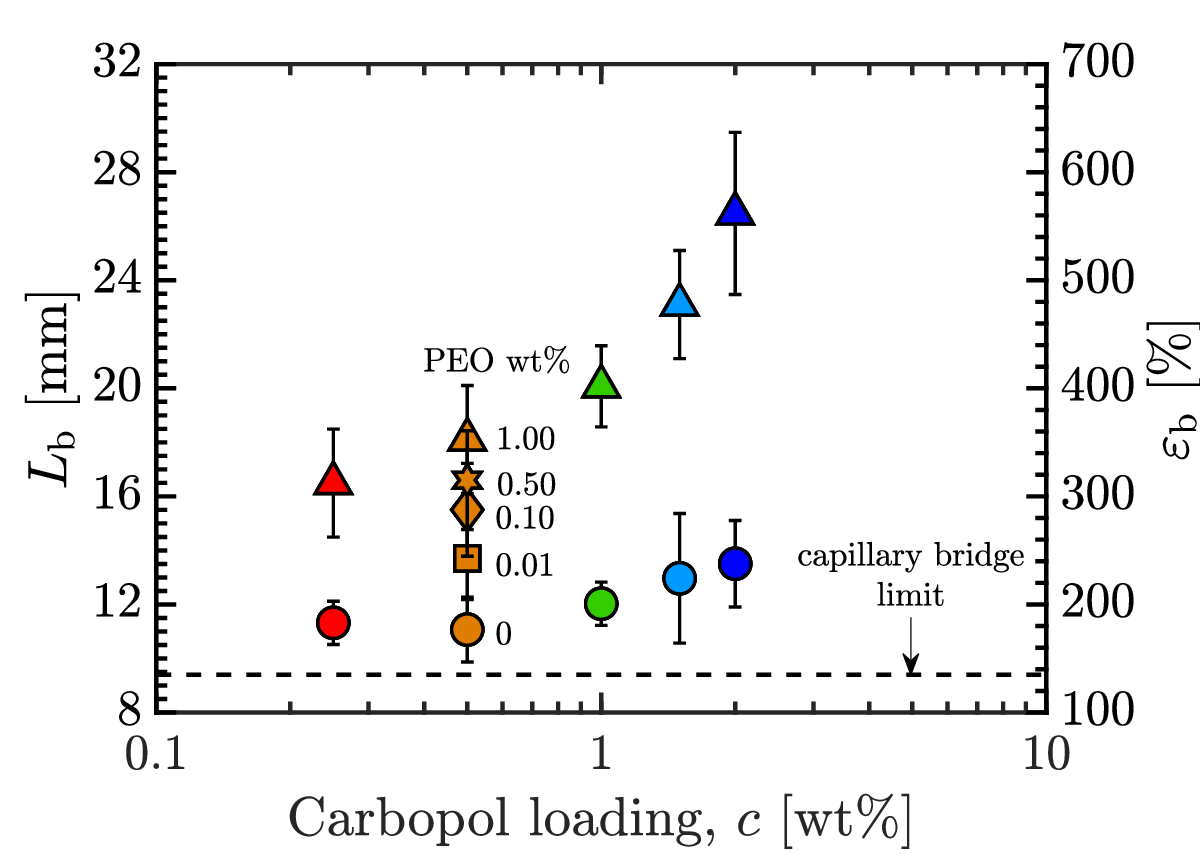}
		\caption{\label{fig:fiser-eSTB}Extensional length-to-break, $L_{\rm b}$, and engineering strain-to-break, $\varepsilon_{\rm b}$, \emph{versus} Carbopol loading, $c$~wt\%. Also shown is the theoretical breakup limit for a capillary bridge of an inviscid fluid  at the same initial aspect ratio used in experiments \cite{Nelson2018}.}
	\end{figure}
	
	The effect of PEO addition on $\varepsilon_{\rm b}$ is consistent across all Carbopol concentrations, as plotted in Fig.~\ref{fig:fiser-eSTB}. We see that $\varepsilon_{\rm b}$ increases with $c$ for both 0 and 1~wt\% PEO. Further, for each value of $c$, adding PEO increases $\varepsilon_{\rm b}$ many fold: $\varepsilon_{\rm b} \leq 250\%$ for all pure Carbopol samples, while $\varepsilon_{\rm b} \geq 350\%$ for all Carbopol+PEO samples with 1~wt\% PEO.
	
	
	
	
	This can be rationalized by considering the effective concentration of PEO chains in the Carbopol suspension and the effect of the Carbopol concentration on the yield stress and viscosity. As is evident in Fig.~\ref{fig:flow-flowcurves}, increasing the Carbopol concentration leads to an increase in both the yield stress and viscosity of the base material. Failure by pinch-off results from the surface tension effects overcoming the resistance imposed by the elastoviscoplastic nature of the material. The dimensionless plastocapillary number, $\mathcal{Y} \equiv \frac{\sigma_{\rm y}}{\Gamma/\ell}$, is defined as the ratio of yield stress to capillary pressure based on a characteristic lengthscale $\ell$ (e.g.\ the filament diameter in our tests); $\Gamma$ is the surface tension of the fluid-air interface \cite{ThompsonSoares2016,JalaalLohse_JFM2019,SSRHE_JFM2020}. The plastocapillary length is defined at $\mathcal{Y} = 1$, such that $\mathscr{L} \equiv \Gamma/\sigma_{\rm y}$. For a given material, when $\ell > \mathscr{L}$, surface tension stresses cannot overcome yield stress and deform the filament. Therefore, materials with a larger yield stress (for similar surface tension values), and thus smaller $\mathscr{L}$, can sustain a stable filament to much smaller lengthscales $\ell$, leading to more stable filaments, and delayed pinch-off. Additionally, since the extensibility of the designed fluid results from the elastic contributions from the dispersed polymer, addition of PEO further increases the extensional strain-to-break. This can be verified by fixing the Carbopol concentration and varying the amount of PEO. We used 0.50~wt\% Carbopol with 0, 0.01, 01, 0.50, and 1.00~wt\% PEO for this test. The data for these has been co-plotted in Fig.~\ref{fig:fiser-eSTB}. We see that $\varepsilon_{\rm b}$ increases monotonically with PEO concentration, which confirms that the increased extensional behavior stems predominantly from the added polymers. Note that the change in $\varepsilon_{\rm b}$ is much larger compared to the change in $\sigma_{\rm y}$, thus there must be effects other than increased yield stress upon polymer addition that increase extensibility.

	The difference in extensibility due to added PEO shows up both in the kinematics and the dynamics. Using the engineering axial stress during filament stretching, $\sigma_{\rm e}$, we can study the effect of added PEO on forces and stresses. This quantity can be calculated using $F_z$ (data shown in Fig.~1, SI) and the initial filament radius, $R_0$, as
	\begin{align}\label{eq:sigma-engg}
		\sigma_{\rm e}(t) = \frac{F_z(t)}{\pi R_0^2}.
	\end{align}
	
\begin{figure}[!ht]
		\begin{minipage}[!ht]{0.49\textwidth}
			\centering
			(a)
			\includegraphics[width=\linewidth]{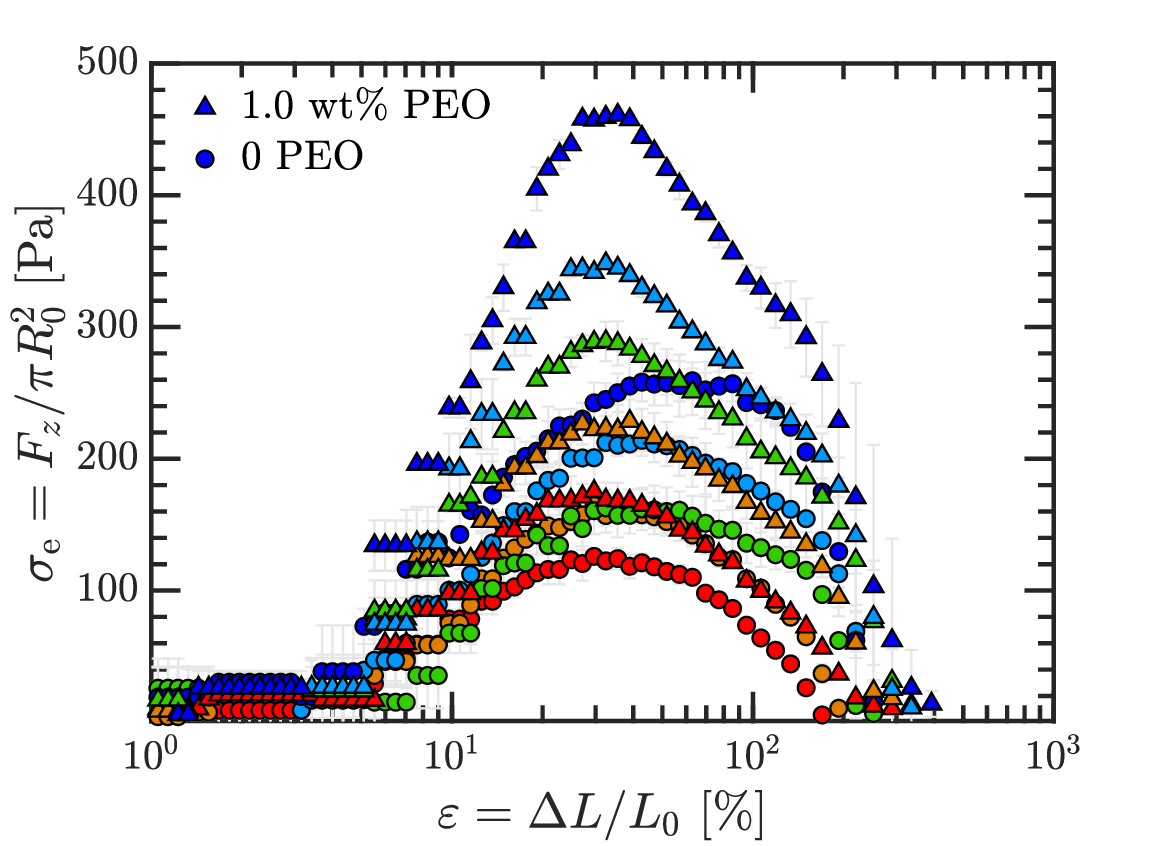}
		\end{minipage}
		\begin{minipage}[!ht]{0.49\textwidth}
			\centering
			(b)
			\includegraphics[width=\linewidth]{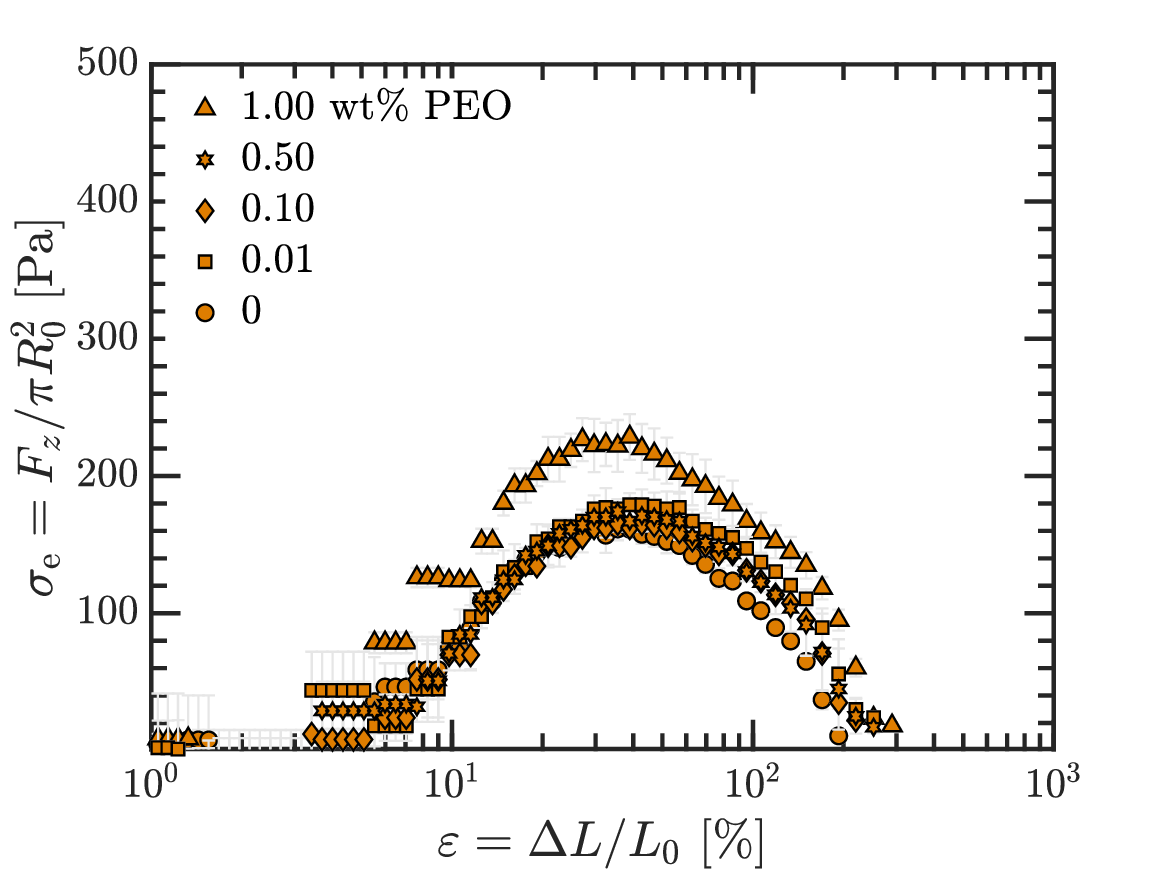}
		\end{minipage}
		\caption{\label{fig:fiser-stress-engg}Engineering axial stress, $\sigma_{\rm e}$, plotted against engineering strain, $\varepsilon$, in filament stretching tests. (a) Data for 0 and 1~wt\% PEO, for all Carbopol concentrations. Carbopol with PEO has a higher $\sigma_{\rm e}$ compared to pure samples for all $c$; for a given Carbopol concentration, $\sigma_{\rm e}$ increases from 0 to 1~wt\% PEO. (b) Data for 0.50~wt\% Carbopol with 0, 0.01, 0.10, 0.50, and 1.00 wt\% PEO added, and $\sigma_{\rm e}$ increases as PEO content increases. The triangles refer to the Carbopol samples with 1~wt\%, and circles to those with 0 PEO.}
	\end{figure}

	The engineering stress data is plotted against engineering strain in Fig.~\ref{fig:fiser-stress-engg}(a). The initial linear increase of the $\sigma_{\rm e}$ \emph{versus} $\varepsilon$ data corresponds to the elastic response of the material at smaller strains. Since we are reporting the engineering stress, and not the true stress \cite{Callister}, a maximum in the engineering stress is reached since the effective area at the region of necking and subsequent failure decreases with deformation. We see that for a given $c$, the maximum engineering stress increases as PEO is added to Carbopol. The ultimate tensile stress, herein interpreted as a measure of yield stress in extension, is defined as
	\begin{align}\label{eq:sigma-U}
		\sigma_{\rm U} \equiv \max_{t>0} \sigma_{\rm e}(t) \equiv \max_{\varepsilon>0} \sigma_{\rm e}(\varepsilon),
	\end{align}
	and is a measure of this maximum stress, which increases with added PEO. This is a signature of increased resistance to extensional deformation.

	Similar to $\varepsilon_{\rm b}$, $\sigma_{\rm U}$ also increases with $c$ for a fixed PEO concentration, which further supports the reasoning that increasingly confined PEO chains in more concentrated Carbopol suspensions lead to greater extensional stresses. In Fig.~\ref{fig:fiser-stress-engg}(b), we have shown $\sigma_{\rm e}$ for 0.50~wt\% Carbopol with 0, 0.01, 01, 0.50, and 1.00~wt\% PEO. We see that $\sigma_{\rm U}$ increases with PEO content, which follows from the reasoning behind similar monotonic trends observed for $\varepsilon_{\rm b}$.

\begin{figure}[!ht]
		\begin{minipage}[!ht]{0.49\textwidth}
			\centering
			(a)
			\includegraphics[width=\linewidth]{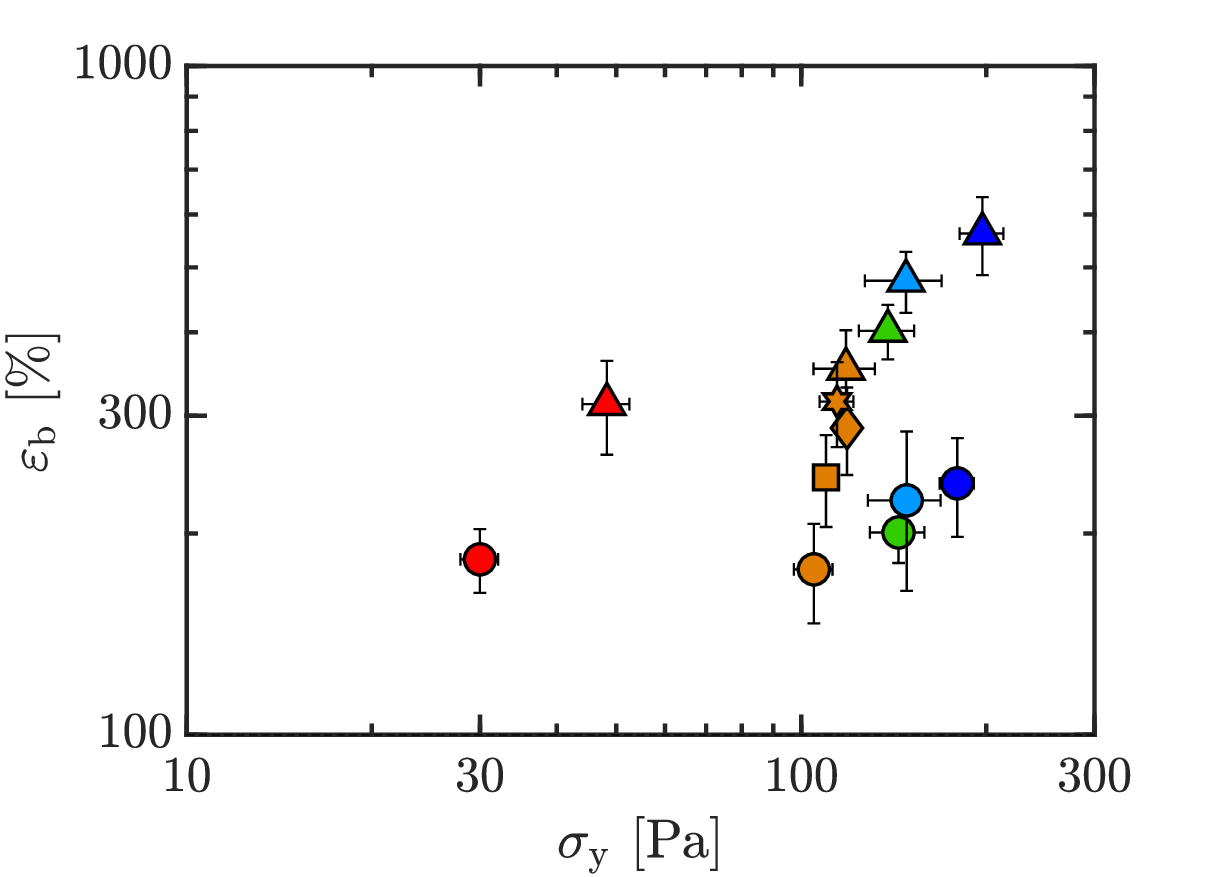}
		\end{minipage}
		\begin{minipage}[!ht]{0.49\textwidth}
			\centering
			(b)
			\includegraphics[width=\linewidth]{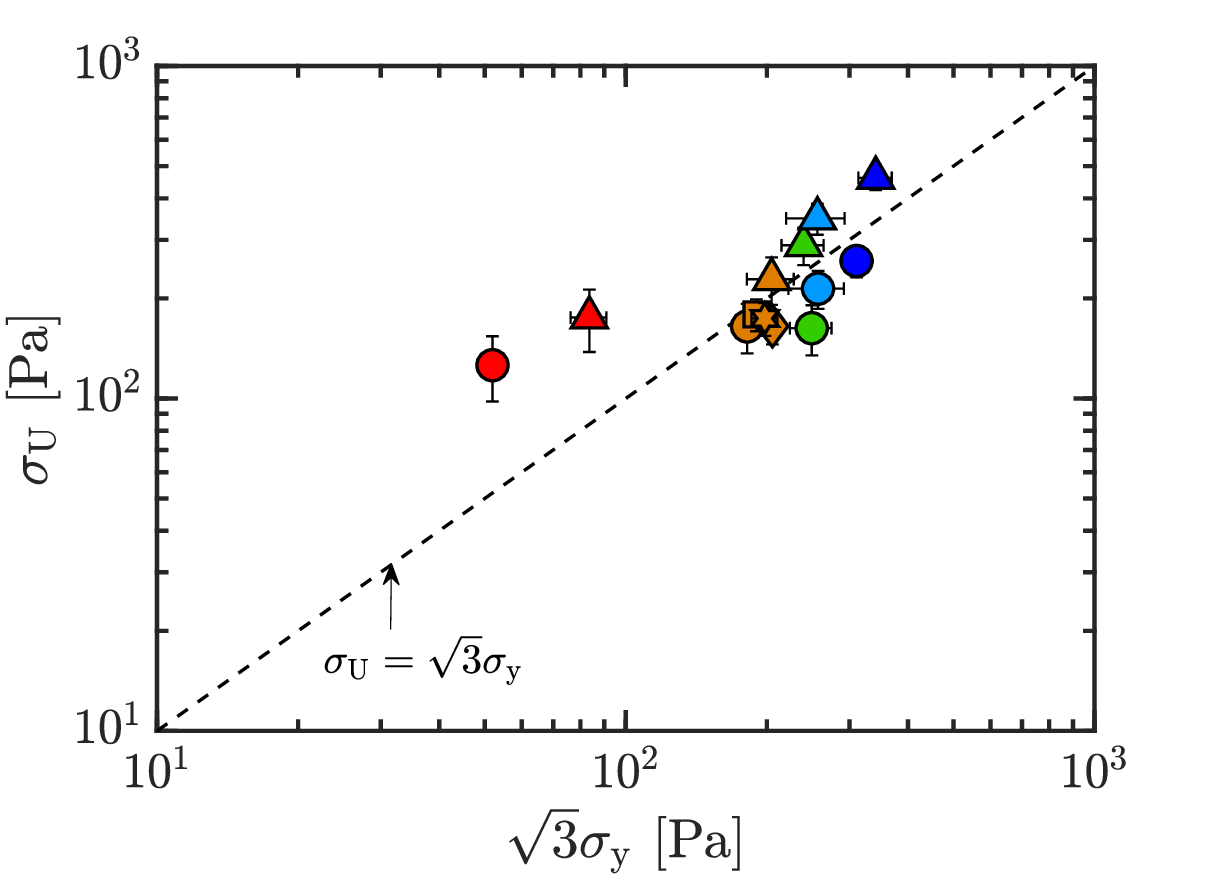}
		\end{minipage}
		\caption{\label{fig:eSTB-eUTS-ys}Ashby co-plots of (a) extensional strain-to-break, $\varepsilon_{\rm b}$, and (b) ultimate extensional stress, $\sigma_{\rm U}$, \emph{versus} shear yield stress $\sigma_{\rm y}$ (from steady shear data), for 0 and 1~wt\% PEO, all wt\% Carbopol. Co-plotted are results for 0.50~wt\% Carbopol with 0, 0.01, 0.10, 0.50, and 1.00 wt\% PEO added. Co-plotted as a dashed line in (b) is the von Mises criterion for yield stresses in shear and extension. The triangles refer to the Carbopol samples with 1~wt\%, and circles to those with 0 PEO. Colors indicate the Carbopol concentration and follow the notation established in Fig. \ref{fig:fiser-eSTB}.}
	\end{figure}

	Based on the data for $\varepsilon_{\rm b}$ and $\sigma_{\rm U}$, we can plot each against $\sigma_{\rm y}$ data from Sec.~\ref{sec:shear} to get Ashby-style co-plots \cite{Ashby_book2011}. In Fig.~\ref{fig:eSTB-eUTS-ys}(a), we show $\varepsilon_{\rm b}$, measured in extension, against $\sigma_{\rm y}$, measured in shear. We see that, for a given $c$, $\varepsilon_{\rm b}$ increases significantly on adding PEO to Carbopol, while the yield stress increases marginally (except for the lowest Carbopol loading of $c=0.25~wt\%$). This is the most clear evidence that it is possible to largely decouple the control of extensibility of this model fluid from its shear properties. Specifically, the extensibility can be increased significantly while suppressing changes in shear properties. Spanning the polymer concentrations between 0 and 1~wt\% PEO for 0.50~wt\% Carbopol changes the strain-to-break monotonically, which also shows that extensibility can be systematically varied whilst superimposing different values of the shear yield stress and extensional failure strains.
	
	In Fig.~\ref{fig:eSTB-eUTS-ys}(b), we plot $\sigma_{\rm U}$, a measure of yield stress in extension, against $\sqrt{3}\sigma_{\rm y}$, a measure of effective yield stress in extension based on shear stresses according to the von Mises criterion \cite{Saramito_JNNFM2007,Saramito_JNNFM2009,Callister}. The dashed line is therefore for materials where the von Mises criterion holds. Pure Carbopol samples fall on or below this line, which means that the ultimate stress is smaller than predicted for ideally elastic materials. Adding PEO to these samples pushes the data above this boundary, which hints at enhanced extensibility, and a fundamentally different microstructural consequence of adding PEO to otherwise largely inextensible yield-stress fluids. Note that the data for 0.25~wt\% Carbopol is an outlier here as well. This might result from the lower value of the yield stress of the material, such that the values of $\sigma_{\rm U}$ measured under extension in this case are more significantly affected by capillary forces, and by experimental noise from the force transducer. The agreement between the yield stresses in shear and extension with the von Mises criterion for the higher concentrations of Carbopol are in line with the observations from the literature \cite{sica2020mises}, where 0.5~wt\% Carbopol suspensions were adequately described by the von Mises yield criterion.
	
	

	This section has elucidated the significant effect of adding polymers to Carbopol suspensions:  steady shear properties, of which the shear yield stress is arguably the most important, were relatively unchanged upon addition of polymer. Additionally, upon PEO addition, shear normal stress increased by a factor of 1.5-2, which is also comparable to the increase in extensibility (2-3 times increase in failure strains) for Carbopol samples with PEO when compared with pure Carbopol. In the following section, we look at the effect of added PEO on viscoelastic properties of Carbopol, and see if the physics of largely decoupled of shear and extension properties is retained.


	\section{\label{sec:shear-osc}Effect of PEO on shear viscoelastic properties}
	In this section, we characterize the oscillatory shear rheology of the ExYSF to study the effect of added PEO on the viscoelastic behavior of Carbopol. Each of these tests is aimed at testing if shear viscoelastic properties are significantly affected by addition of the polymer compared to the change in extensional behavior. We show frequency sweeps done in the linear regime (small amplitude oscillatory shear or SAOS) and oscillatory strain amplitude sweeps spanning linear to nonlinear regimes (large amplitude oscillatory shear strain or LAOStrain) .
	
	\subsection{\label{subsec:SAOS}Frequency sweeps in SAOS}
\begin{figure}[!ht]
		\begin{minipage}[!ht]{0.49\textwidth}
			\centering
			(a)
			\includegraphics[width=\linewidth]{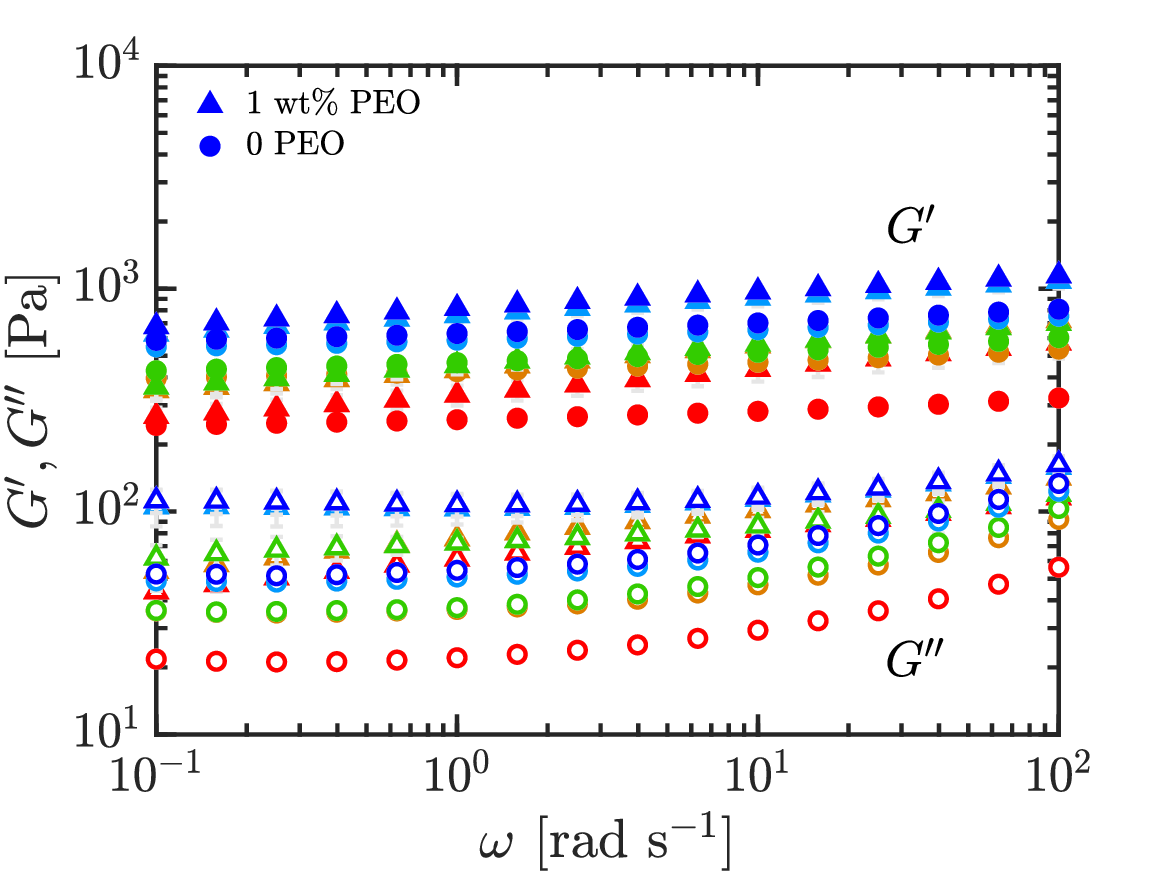}
		\end{minipage}
		\begin{minipage}[!ht]{0.49\textwidth}
			\centering
			(b)
			\includegraphics[width=\linewidth]{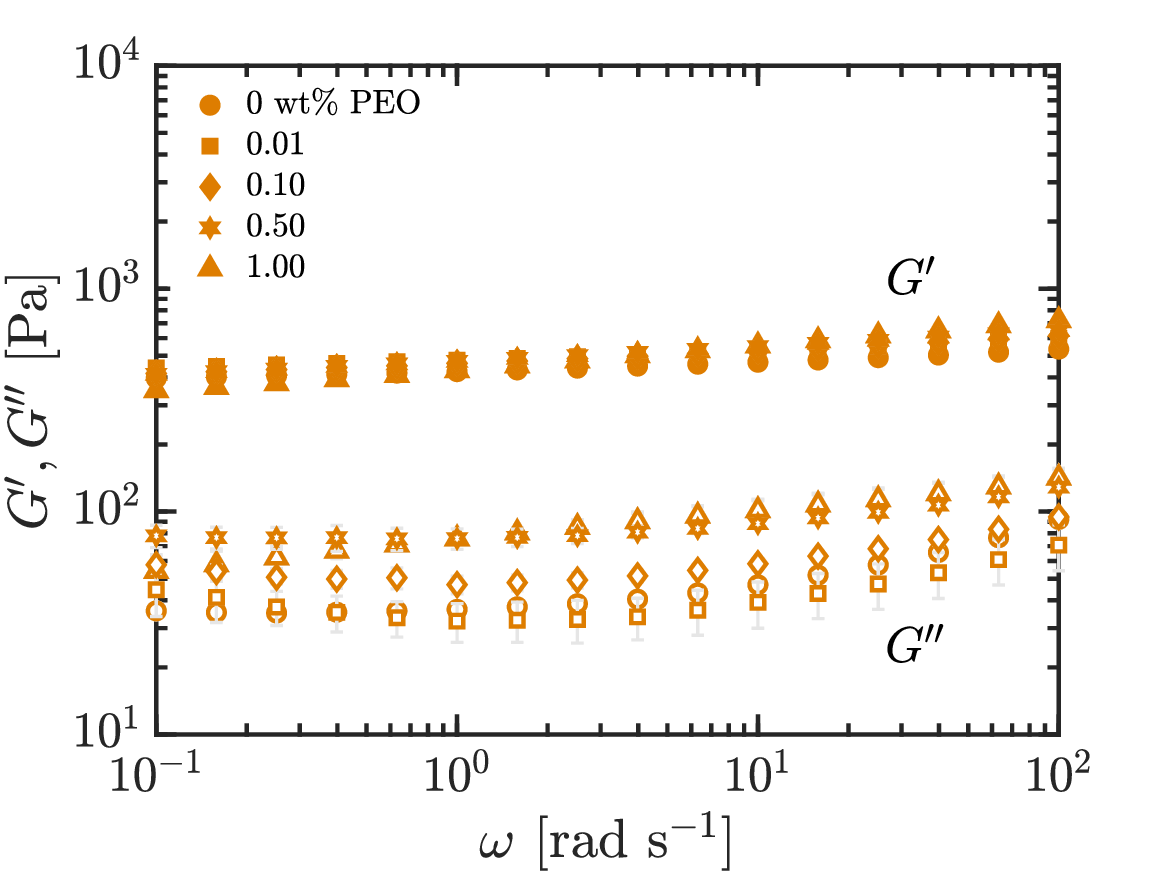}
		\end{minipage}
		\caption{\label{fig:SAOS-Gvfreq}Comparison of viscoelastic moduli in oscillatory shear frequency sweeps plotted as a function of frequency for different Carbopol+PEO samples; $\gamma_0 = 1\%$ and $\omega = $~0.1-100~rad~s$^{-1}$. (a) All wt\% Carbopol with 0 or 1~wt\% PEO added. (b) 0.50~wt\% Carbopol with 0, 0.01, 0.10, 0.50, and 1.00 wt\% PEO added. The triangles refer to the Carbopol samples with 1~wt\%, and circles to those with 0 PEO.}
	\end{figure}

	Frequency sweeps were done at a constant strain amplitude of $\gamma_0=1\%$, which was in the linear regime as determined from the strain amplitude sweeps (shown next). The frequency was varied from $\omega = 0.1$-$100~{\rm rad~s}^{-1}$, and the results are shown in Fig.~\ref{fig:SAOS-Gvfreq}.

	Two features of the plots stand out: firstly, the values of $G^{\prime}$ are greater than $G^{\prime\prime}$; $\tan\delta = G^{\prime\prime}/G^{\prime} < 0.1$ for all pure Carbopol samples. This indicates that the material has a predominantly elastic response, which is  characteristic of viscoelastic solids such as gels, arrested systems, crowded and jammed suspensions, and soft glassy solids \cite{CoussotReview2014,BonnManneville2017,Carbopol_Ovarlez}. Being a soft glassy system, Carbopol is expected to display this behavior below the yield stress \cite{Piau_cpol2007,SethBonnecaze_NatMat2011,Carbopol_Ovarlez}. This primary signature is not changed by the addition of PEO, although now we have $\tan\delta < 0.2$, indicating increased dissipative effects due to added PEO (see Fig.~\ref{fig:Gp0-tandelta-conc}).
	
	Secondly, $G^{\prime}$ is only weakly dependent on frequency (average log-log slope $\ln G^\prime / \ln \omega < 0.1$). This has also been reported previously for Carbopol and other microgels \cite{Piau_cpol2007,SethBonnecaze_NatMat2011,AsheshGaurav_SM2019}. This is characteristic of yield stress fluids and condensed matter in general, whether gels or soft glasses \cite{BonnManneville2017}, and signifies extremely slow thermal relaxation of the microstructure (and thus a largely constant elastic modulus over a wide range of frequencies) due to energetically arrested particulate networks or glassy dynamics. This characteristic is also unchanged upon adding PEO to Carbopol. We also see that $G^{\prime\prime}$ increases with frequency, for both pure Carbopol and Carbopol+PEO, suggesting that over shorter timescales the material displays a more pronounced energy dissipation.
	
	More detailed viscoelastic data is shown by creep tests in the SI, which show these materials are predominantly solid-like, for stresses below the yield stress, for timescales up to at least 600~s. This duration is longer than 10~s (corresponding to $\omega = 0.1~{\rm rad~s}^{-1}$), which is the longest accessible timescale from the SAOS data shown here. The addition of PEO does not significantly change the shapes of the viscoelastic creep compliance curves.
	


	\subsection{\label{subsec:LAOS}Strain sweeps in LAOStrain}
\begin{figure}[!ht]
		\begin{minipage}[!ht]{0.49\textwidth}
			\centering
			(a)
			\includegraphics[width=\linewidth]{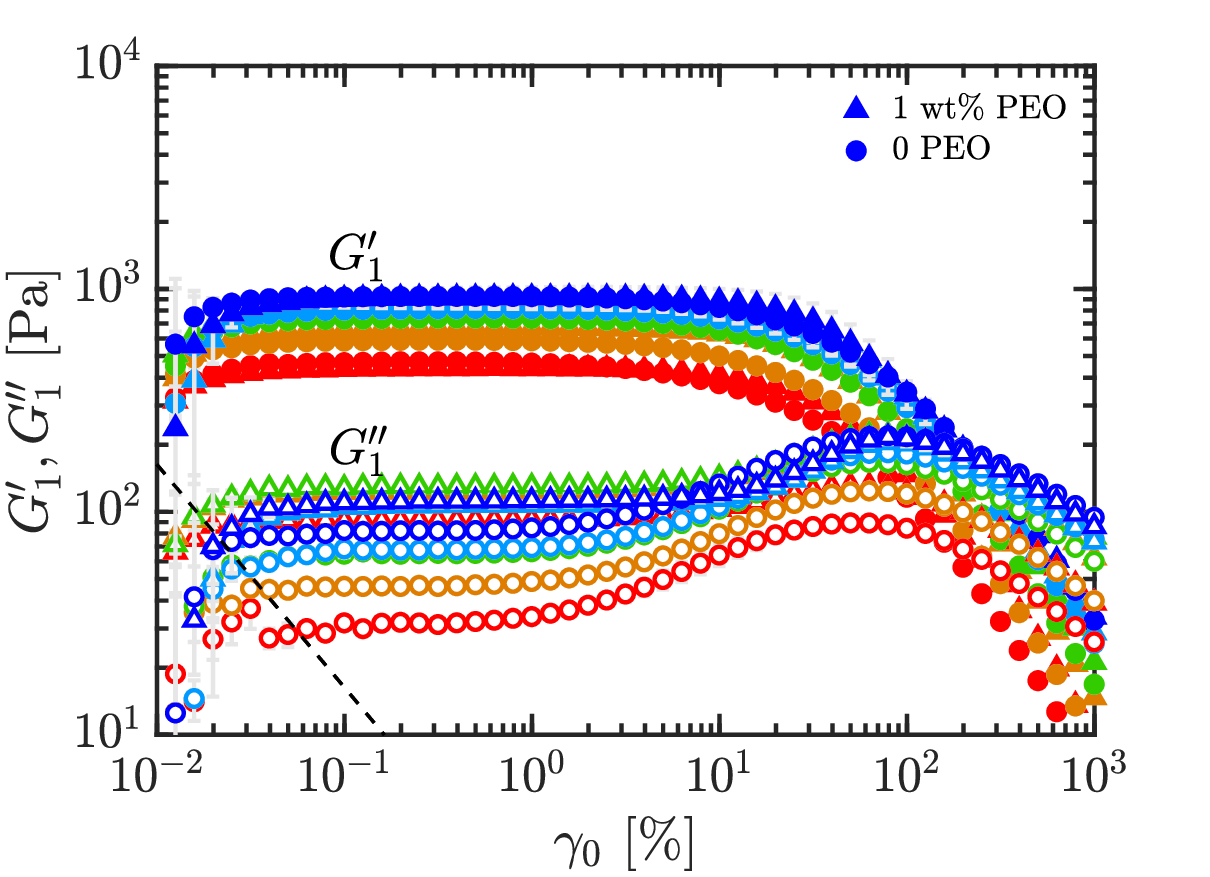}
		\end{minipage}
		\begin{minipage}[!ht]{0.49\textwidth}
			\centering
			(b)
			\includegraphics[width=\linewidth]{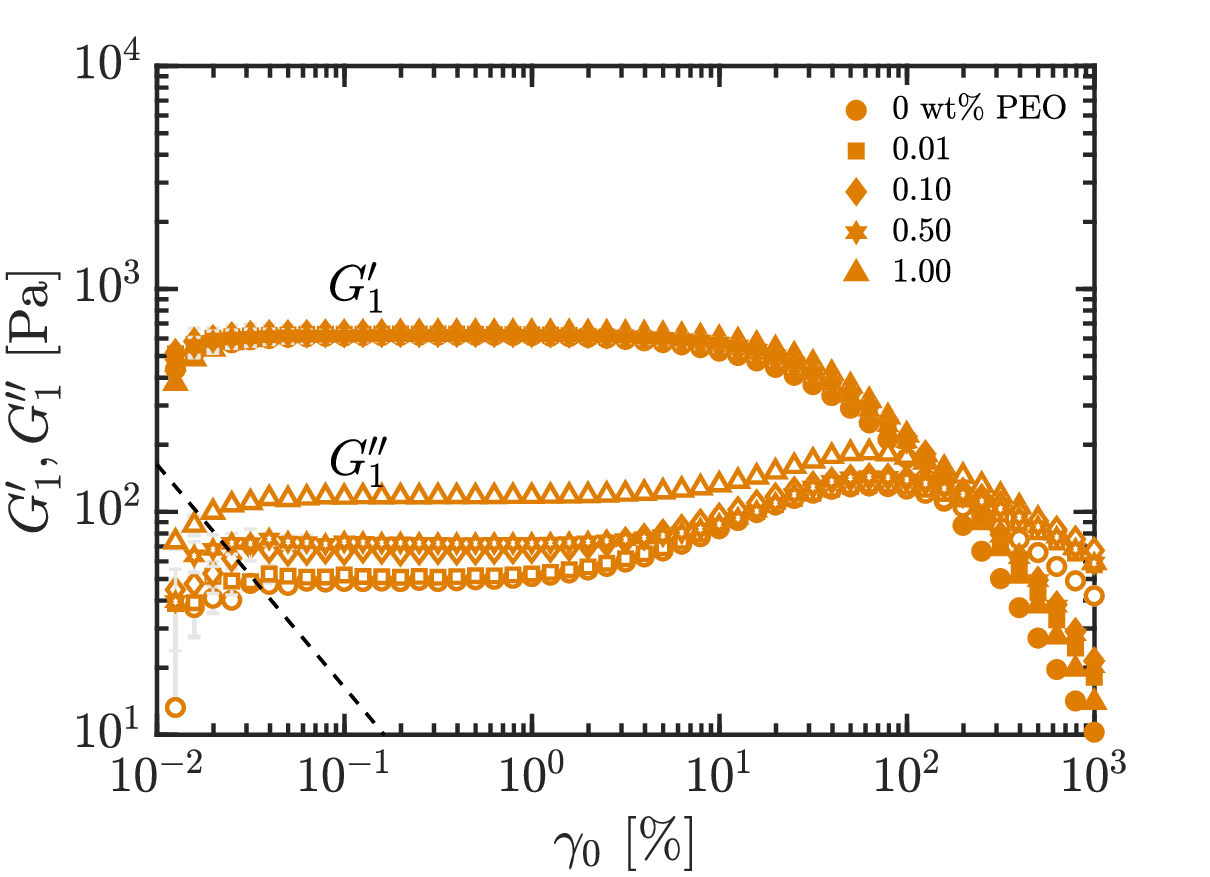}
		\end{minipage}
		\caption{\label{fig:LAOS-Gvstrain}Comparison of first harmonic storage and loss viscoelastic moduli in oscillatory shear strain sweeps, plotted as a function of the strain amplitude during the cycle, for different Carbopol+PEO samples; $\omega=10$~rad~s$^{-1}$ and $\gamma_0 = $~0.01-1000\%. (a) All wt\% Carbopol with 0 or 1~wt\% PEO added. (b) 0.50~wt\% Carbopol with 0, 0.01, 0.10, 0.50, and 1.00 wt\% PEO added. The dashed line in each plot shows the low-torque limit of the instrument \cite{RHE_baddata}, below which the data cannot be trusted. The triangles refer to the Carbopol samples with 1~wt\% PEO, and circles to those with 0 PEO.}
	\end{figure}

	In strain sweeps, the oscillatory strain amplitude was varied from $\gamma_0 = 0.01$-$1000\%$. All tests were conducted at a fixed frequency of $\omega=10~{\rm rad~s}^{-1}$. The results of the strain sweeps are shown in Fig.~\ref{fig:LAOS-Gvstrain}, with the viscoelastic moduli plotted as a function of the oscillatory strain amplitude, $\gamma_0$. Here, we report the first harmonic storage $\left(G^{\prime}_1\right)$ and loss $\left(G^{\prime\prime}_1\right)$ moduli \cite{RHE_JoR2013}.
	
	In the nonlinear viscoelastic regime, higher-order harmonics distort the periodic stress response \cite{Ewoldt2008}. For both pure Carbopol and Carbopol+PEO, we see that $G^{\prime}_1$ is independent of $\gamma_0$ at low amplitudes, indicating a response within the linear viscoelastic regime. At higher strains, the modulus rolls-off and decreases dramatically, displaying a softening behavior that is common for colloidal gels and soft glasses \cite{Piau_cpol2007,CoussotReview2014,Carbopol_Cates,SethBonnecaze_NatMat2011,Carbopol_Kim,Carbopol_Ovarlez,BonnManneville2017}. The qualitative behavior is independent of PEO concentration. Similarly, $G^{\prime\prime}_1$ also shows a response within the linear viscoelastic regime at low strain amplitudes. At intermediate strain amplitudes, an overshoot is observed, which is characteristic of Type III behavior in LAOS as defined by Hyun \emph{et al.} \cite{Hyun2002}. This suggests that the material displays increased dissipative processes, which have been recently explained by plastic deformations present at moderate strain amplitudes \cite{Donley_PNAS2020} in glassy systems. $G^{\prime\prime}_1$ eventually rolls-off at larger strain amplitudes. The critical strain at the departure from linear viscoelasticity does not seem to be significantly affected by addition of PEO, since the deviation from linearity in $G^\prime_1$ happens around $\gamma_{0}\simeq 10\%$ for all samples. The same features are observed upon addition of PEO.
	
	Fig.~\ref{fig:LAOS-Gvstrain}(b) shows the moduli when different amounts of PEO are added to the 0.50~wt\% Carbopol sample. No significant changes are evident on the values of $G^{\prime}_1$, suggesting that the elastic behavior of the sample is largely agnostic to the presence of the polymer. This surprising result can be understood as the linear elastic behavior of the material arising primarily from the jammed microstructure created by the Carbopol microgel particles, which is not changed when the polymer is added. A slight increase in $G^{\prime\prime}_1$ can be seen (roughly half an order of magnitude) upon increasing the amount of PEO. This again suggests that addition of the polymer increases the energy dissipation in the material, but it is not possible to identify if this increase in energy dissipation is due to plastic or viscous deformations.

	\subsection{Summarizing the effect of PEO on LVE of Carbopol}\label{subsec:LVE-summary}
\begin{figure}[!ht]
		\begin{minipage}[!ht]{0.49\textwidth}
			\centering
			(a)
			\includegraphics[width=\linewidth]{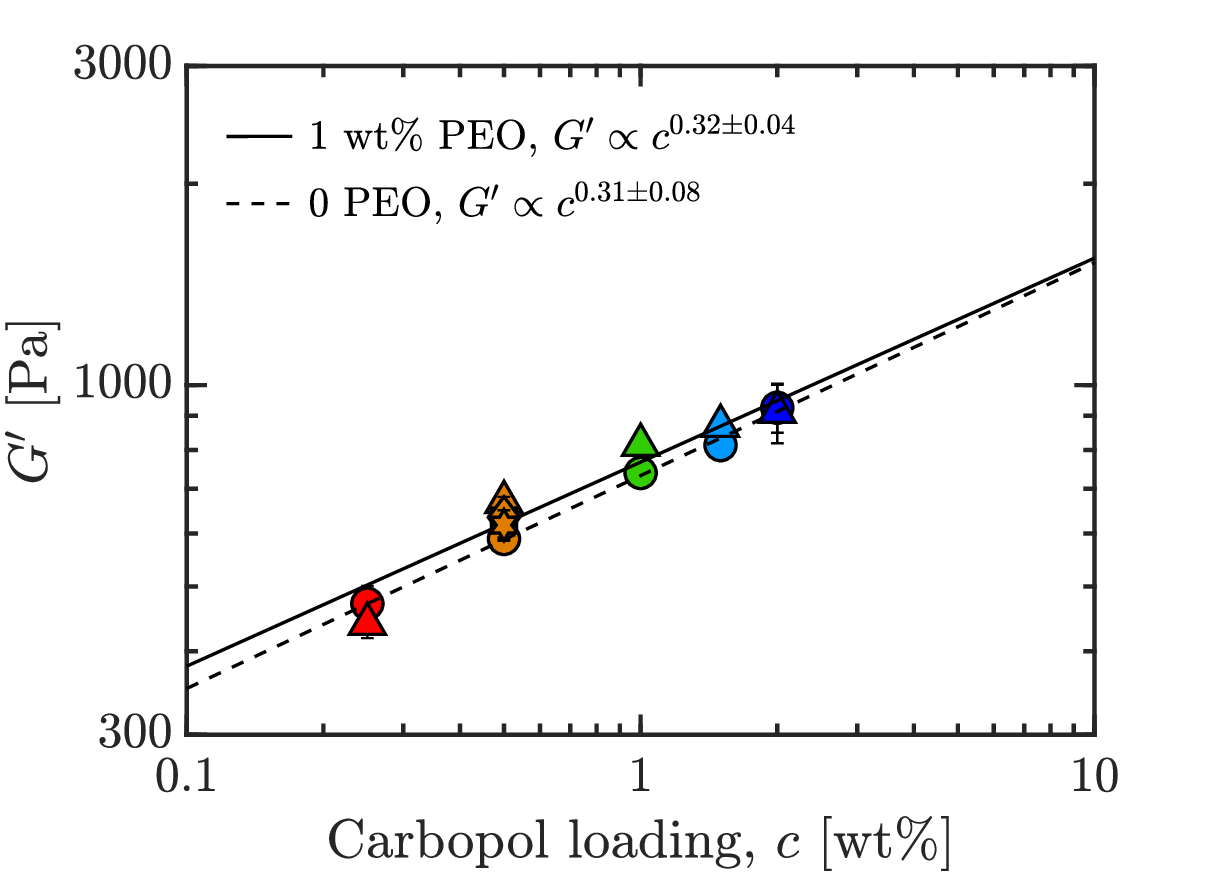}
		\end{minipage}
		\begin{minipage}[!ht]{0.49\textwidth}
			\centering
			(b)
			\includegraphics[width=\linewidth]{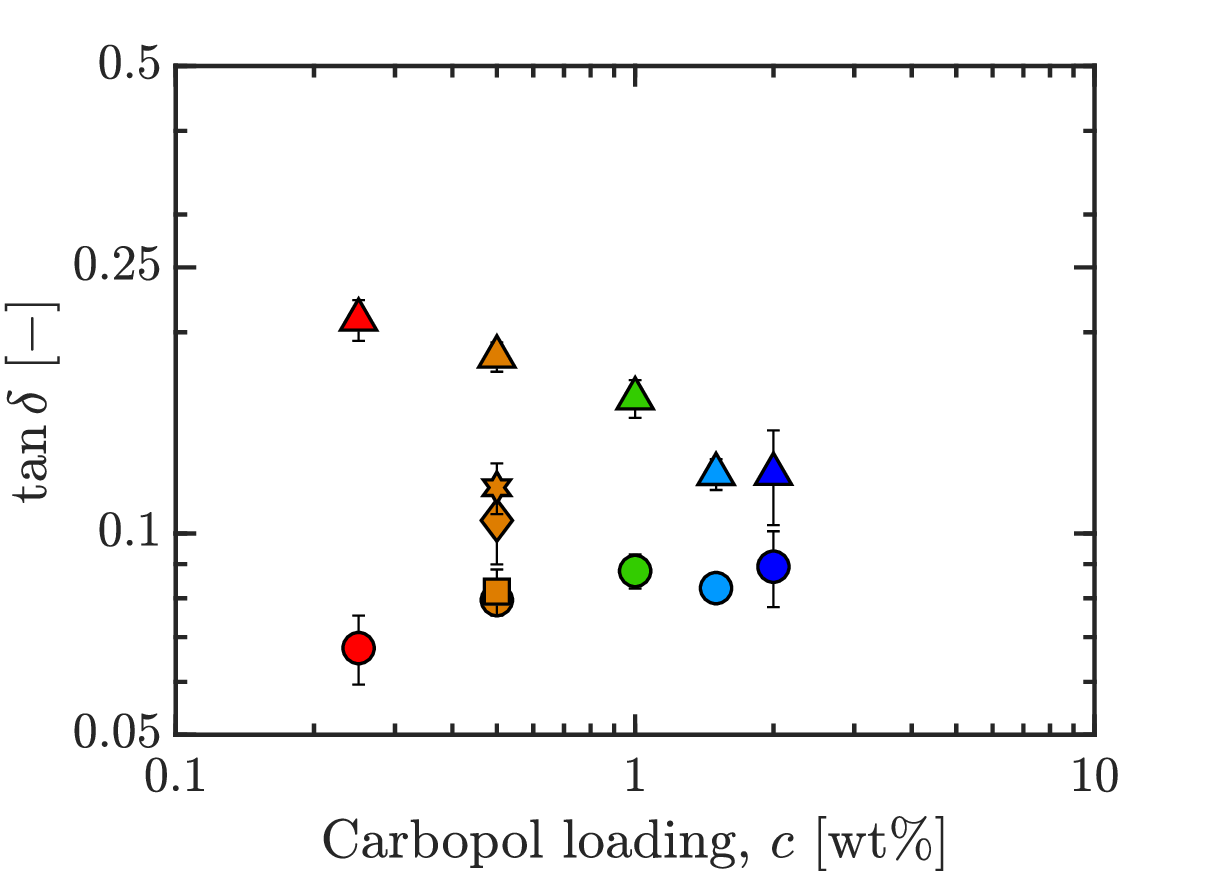}
		\end{minipage}
		\caption{\label{fig:Gp0-tandelta-conc}Co-plots of (a) linear storage modulus $G^\prime$, and (b) $\tan\delta \equiv G^{\prime\prime}/G^\prime$, from LAOStrain, at $\omega = 10~{\rm rad}~{\rm s}^{-1}$, \emph{versus} Carbopol loading, $c$, for 0 and 1~wt\% PEO, all wt\% Carbopol. Co-plotted are results for 0.50~wt\% Carbopol with 0, 0.01, 0.10, 0.50, and 1.00 wt\% PEO added. The fit lines in (a) are power-laws for $G^\prime$ as function of $c$, and the uncertainties on the scaling exponents are from data and fit uncertainties. The triangles refer to the Carbopol samples with 1~wt\%, and circles to those with 0 PEO.}
	\end{figure}

	The effect of PEO on the linear viscoelastic properties of Carbopol can be more clearly seen from the linear regime in the LAOStrain data. We can look at the linear storage modulus at a fixed frequency calculated from LAOS data, as $G^\prime = \lim_{\gamma_0 \rightarrow 0}G^\prime_1$, and the ratio of the loss to storage moduli, $\tan\delta \equiv G^{\prime\prime}/G^\prime$, where $G^{\prime\prime} = \lim_{\gamma_0 \rightarrow 0}G^{\prime\prime}_1$. In Fig.~\ref{fig:Gp0-tandelta-conc}(a), we plot the power law scaling of $G^\prime$ with Carbopol loading $c$. We see that the scaling exponent in $G^\prime \propto c^\nu$ is $\nu = 0.31 \pm 0.08$ for pure Carbopol, which is very similar to the exponent calculated upon addition of PEO, for which $\nu = 0.32 \pm 0.04$. These values of $\nu$ are close to those reported in the literature for Carbopol \cite{Piau_cpol2007,AZN_SM2017}. The magnitudes of $G^\prime$ are also very close for Carbopol samples with and without PEO.
	
	In Fig.~\ref{fig:Gp0-tandelta-conc}(b), we see that $\tan\delta$ in the linear viscoelastic regime is much larger with PEO than without. This comes from a more significant increase in the $G^{\prime\prime}$ compared to $G^\prime$ with polymer addition, as can be seen in Fig.~\ref{fig:LAOS-Gvstrain}(b). The higher value of the loss modulus upon addition of PEO suggests increased dissipation, as expected due to the increased amounts of polymer in the suspension. So, although the dissipative component of the stress increases with polymer addition, the storage modulus is largely unaffected in both its magnitude and scaling with $c$, since the Carbopol particles of the ExYSF contribute to the bulk of the stiffness in shear. These results are in agreement with those seen for the yield stress from the steady shear data, i.e.\ the increase in Carbopol concentration increases the yield stress measured in the steady-state flow curves, and a less significant effect is observed when PEO is added to the samples.

	\begin{figure}[!ht]
		\centering
		\includegraphics[width=0.5\linewidth]{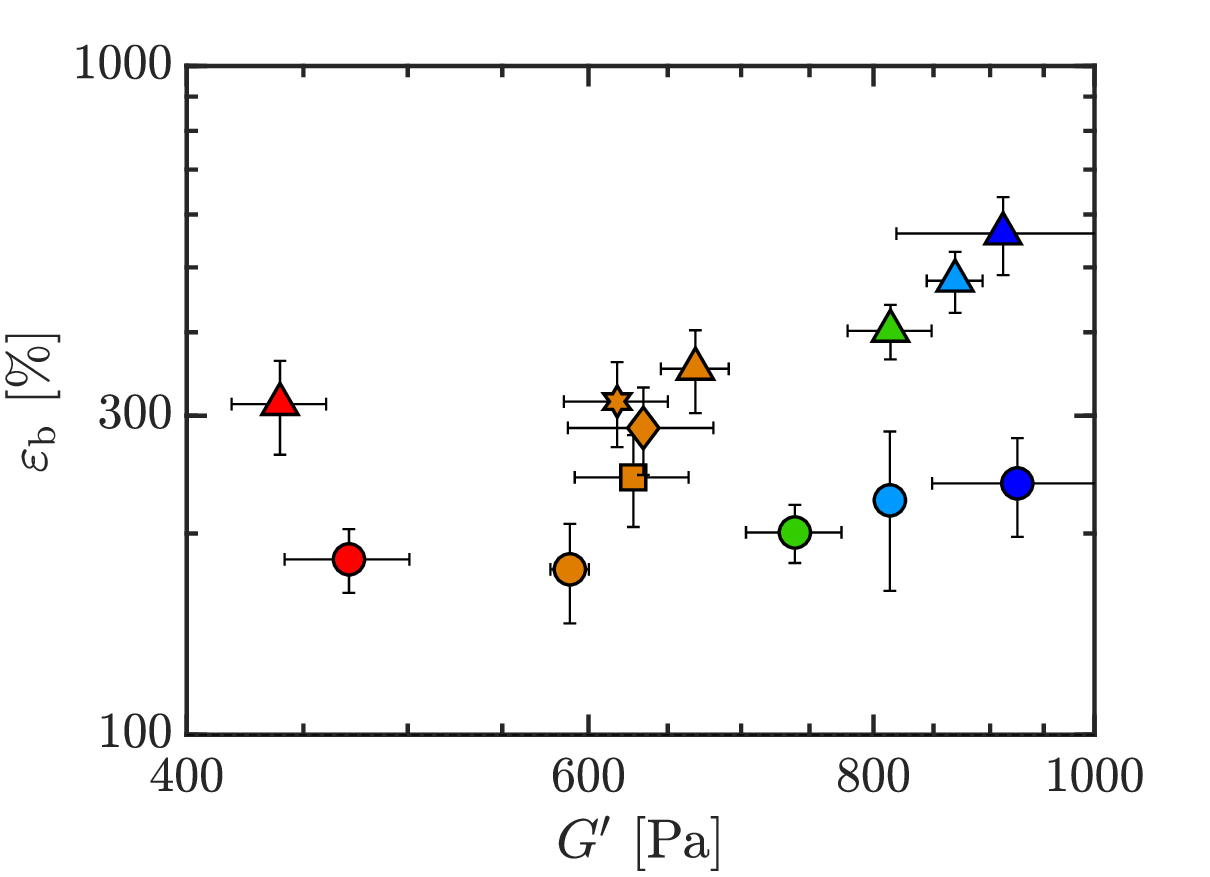}
		\caption{\label{fig:eSTB-params-Gp0}Ashby co-plots of extensional strain-to-break \emph{versus} linear storage modulus from LAOS, $G^\prime$ at $\omega = 10~{\rm rad}~{\rm s}^{-1}$, for 0 and 1~wt\% PEO, all wt\% Carbopol. Co-plotted are results for 0.50~wt\% Carbopol with 0, 0.01, 0.10, 0.50, and 1.00 wt\% PEO added. The triangles refer to the Carbopol samples with 1~wt\% PEO, and circles to those with 0 PEO.}
	\end{figure}

	Based on the LAOS data, we can compare the metrics in extension to those in shear viscoelasticity. Fig.~\ref{fig:eSTB-params-Gp0} plots $\varepsilon_{\rm b}$ against $G^\prime$, and we see very similar trends as with yield stress, Fig.~\ref{fig:eSTB-eUTS-ys}(a). The values of $\varepsilon_{\rm b}$ increase with added PEO for all Carbopol concentrations, while $G^\prime$ is mostly unaltered upon adding PEO. The increase in extensibility with PEO content while maintaining similar $G^\prime$ is systematically tunable, as seen from the intermediate PEO concentrations for 0.50~wt\% Carbopol data (yellow symbols). We therefore see that both steady shear properties such as $\sigma_{\rm y}$ as well as linear viscoelastic properties such as $G^\prime$ are marginally changed when PEO is added, and the extensibility can be increased almost independently from the change in shear properties for this model fluid.

	
	\section{\label{sec:discussion}Discussion}
\begin{figure}[!ht]
		\centering
		\includegraphics[width=\linewidth]{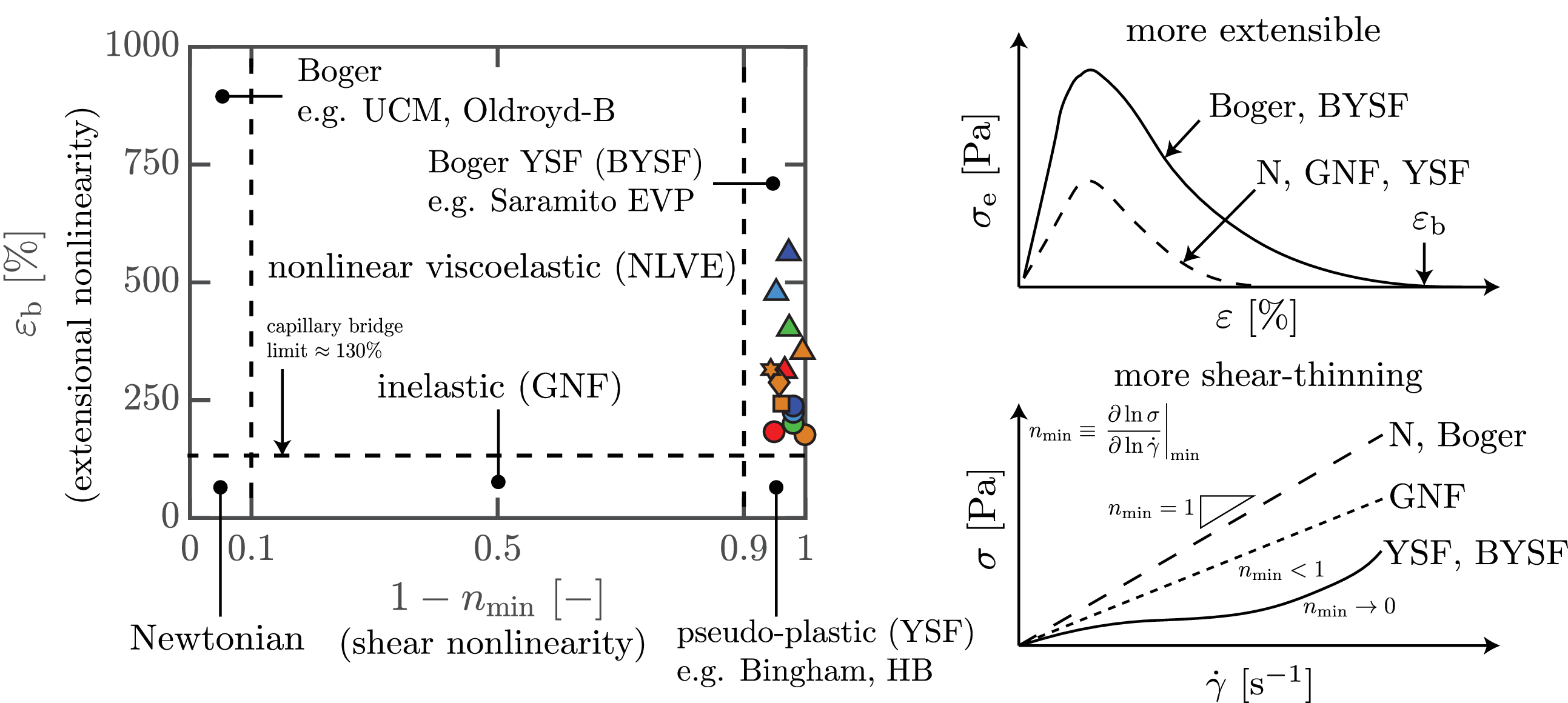}
		\caption{\label{fig:Ashby-Boger}Ashby diagram of nonlinearities in shear against extension, with reference to Boger Fluids \cite{Boger-Boger,JamesARFM2009}. The map helps visualize the different regimes of shear and extensional nonlinearities shown by materials and constitutive models. The colored symbols represent the data for all materials tested in this paper.}
	\end{figure}

	The data presented in this paper lies in a rather unique regime of rheological behavior. In Fig.~\ref{fig:Ashby-Boger}, we propose a map to visualize the amounts of shear \emph{versus} extensional nonlinearities in a material. Different metrics could be used to represent the key rheological features in shear and in extension, so the choice of which metrics to use in this diagram is open to debate. For extensional nonlinearity, here we have used $\varepsilon_{\rm b}$ as a metric, while $1-n_{\rm min}$ is a surrogate for nonlinear behavior in shear, obtained from the steady shear data. The value of $n_{\rm min}$ is defined as the smallest gradient of the steady-state flow curve (see Fig.~\ref{fig:Ashby-Boger} for a schematic), as
	\begin{align}
		n_{\rm min} \equiv \left. \frac{\partial \ln\sigma}{\partial \ln\dot{\gamma}} \right|_{\rm min},
	\end{align}
	such that smaller values of $n_{\rm min}$ give the most extreme regime of shear thinning; $1-n_{\rm min}$ correspondingly approaches 1 as shear nonlinearities become more significant, and approaches zero in the limit of constant shear viscosity. Materials and models that largely follow a constant viscosity in shear, such as the Newtonian fluid model and Boger fluids, are expected to have a value of $n_{\rm min} \rightarrow 1$ (see the bottom-right inset in Fig.~\ref{fig:Ashby-Boger}). Generalized Newtonian fluid models that describe shear thinning behavior, such as the power-law model when $n<1$, are expected to show a variable gradient of the $\sigma$-$\dot{\gamma}$ curve, such that $n_{\rm min}$ is located at large shear rates. Finally, $n_{\rm min}$ is expected to be located within the range of stresses where the yielding transition takes place for yield stress fluids (YSF), since in this region a dramatic increase of shear rates with a small increase of shear stress. It is important to highlight that $n_{\rm min}$ therefore does not immediately distinguish viscoplastic yield-stress fluids from the more general class of  shear-thinning behavior, as it does not take into account the range of shear rates over which $n_{\rm min}$ occurs, or the value of the yield stress. However, all yield stress fluids will have a very low value of $n_{\rm min}$. We see that all our data lie on the right side of the plot, meaning that we have systematically explored varying ranges of extensional nonlinearity for materials with significant nonlinear behavior in shear.
	
	The left edge of the map has a name: Boger fluids \cite{DPL_vol1,Macosko:1994}, a class of rheological response with strong nonlinearity in elastic effects as well as in extension, and minimal nonlinearity of the shear stress. The right edge of the map is the limit of maximum shear stress nonlinearity in steady flow, $n_{\rm min} \rightarrow 0$, the limit of a yield stress fluid, which may also have a range of extensibility. The model material studied here lies along this edge. 
	
	Among the many possible names for this right edge region, we choose the name ``Boger yield-stress fluids'' to evoke the connection to Boger fluids that occupy the opposite edge of behavior. The simplest elastoviscoplastic (EVP) models, such as the Saramito model \cite{Saramito_JNNFM2007, Saramito_JNNFM2009}, also show signatures reminiscent of Boger yield-stress fluids: yield-stress fluids for which the extensional behavior can be tuned independently while maintaining the same shear thinning viscosity exponent.

	
	We have shown that it is possible to formulate samples with various ranges of yield stress and extensibility, and these two features can be decoupled to a large extent. However, for certain concentrations of Carbopol, the two effects are more closely coupled; e.g.\ for 0.25~wt\% Carbopol systems, $\sigma_{\rm y}$ does not follow the general scaling with Carbopol loading $c$, and this hints at two different microstructural regimes of yielding behavior, which has already been discussed earlier. Investigating the effect of added polymers to microgels in this low Carbopol concentration and the extent of decoupling between shear and extensional properties is an important topic for future work.
	
	It must be noted that we have not studied the aging and thixotropy of Carbopol samples with or without PEO. Carbopol is known to show varied degrees of thixotropy and aging \cite{Moller2009,Gordon2017,Ong2019,SSJOR2022,Sen_PhDThesis}, although the degree of structural breakdown or aging has been shown to not be significant \cite{SSJOR2022}. We have thus not tested the thixotropic and aging response of our materials.

	\section{\label{sec:conclusion}Conclusion}
	In this work, we have formulated a model yield-stress fluid with tunable extensional properties. The extensibility is largely decoupled from the shear properties when doping the base Carbopol microgel particle suspensions with high molecular weight PEO. Adding PEO to Carbopol does not change key rheological properties measured in shear experiments, namely the shear yield stress $\sigma_{\rm y}$ and the linear viscoelastic storage modulus $G^\prime$, by more than 10-20\%, so the elastic nature of Carbopol is not changed much, although the dissipation is increased on adding PEO. On the contrary, properties in extension, such as the extensional strain-to-break $\varepsilon_{\rm b}$, are increased many fold. These results suggest that the addition of PEO does not significantly alter the interactions between Carbopol microgel particles at high loadings, which are responsible for the elastoplastic shear behavior, while the extensional properties are enhanced by the PEO chains, which may be largely limited to the solvent phase. This also exemplifies that it is possible to formulate simple mixtures of complex fluids, approximately superimposing their individual properties to obtain a hybrid material with tunable behavior. While there is some variation of shear properties when varying the extensional behavior of this material system, we consider it minimal compared to multi-component systems in general where significant non-trivial coupling may occur. As such, we consider this Carbopol+PEO system a model fluid which can be used to experimentally study the isolated effect of changing extensibility in yield stress fluids, the fundamental understanding of which will benefit applications ranging from spray coating of fire suppressants \cite{Appel_PNAS2016,SSRHE_DFD2019,SSRHE_JFM2020,Sen_PhDThesis} to direct-ink writing additive manufacturing \cite{Rauzan2018,Poole_SM2019}, disease control \cite{disease_Zayas,disease_Stone}, and food texture \cite{texture_Stokes,texture_Nicolas,texture_Fischer}, among many others.
	


\section{Acknowledgments}
This work was funded by the National Science Foundation, CAREER Award, CBET-1351342. The authors thank Dr.\ C.\ Saengow and Dr.\ Y.\ Wang for helpful discussions.

\bibliographystyle{unsrt}
\bibliography{Main}

\begin{thebibliography}{10}

\bibitem{Nelson2018}
A.Z. Nelson, R.E. Bras, J.~Liu, and R.H. Ewoldt.
\newblock {Extending yield-stress fluid paradigms}.
\newblock {\em J. Rheol.}, 62(1):357--369, 2018.

\bibitem{AZN_CurrOpin2019}
A.~Z. Nelson, K.~S. Schweizer, B.~M. Rauzan, R.~G. Nuzzo, J.~Vermant, and R.~H.
  Ewoldt.
\newblock Designing and transforming yield-stress fluids.
\newblock {\em Curr. Opin. Solid State Mater. Sci.}, 23(5):100758, 2019.

\bibitem{Bonn2022}
R.~I. Dekker, H.~V.~M. Kibbelaar, A.~Deblais, and D.~Bonn.
\newblock Rheology of emulsions with polymer solutions as the continuous phase.
\newblock {\em J. Non-Newtonian Fluid Mech.}, 310:104938, 2022.

\bibitem{Piau_cpol2007}
J.~M. Piau.
\newblock {Carbopol gels: Elastoviscoplastic and slippery glasses made of
  individual swollen sponges Meso- and macroscopic properties, constitutive
  equations and scaling laws}.
\newblock {\em J. Non-Newtonian Fluid Mech.}, 144:1--29, 2007.

\bibitem{CoussotReview2014}
P.~Coussot.
\newblock Yield stress fluid flows: A review of experimental data.
\newblock {\em J. Non-Newtonian Fluid Mech.}, 211:31--49, 2014.

\bibitem{Bonn_YS_JNNFM2016}
M.~Dinkgreve, J.~Paredes, M.~M. Denn, and D.~Bonn.
\newblock On different ways of measuring ``the'' yield stress.
\newblock {\em J. Non-Newtonian Fluid Mech.}, 238:233--241, 2016.

\bibitem{BonnManneville2017}
D.~Bonn, M.~M. Denn, L.~Berthier, T.~Divoux, and S.~Manneville.
\newblock Yield stress materials in soft condensed matter.
\newblock {\em Rev. Mod. Phys.}, 89(3):035005, 2017.

\bibitem{sollich1998rheological}
P.~Sollich.
\newblock Rheological constitutive equation for a model of soft glassy
  materials.
\newblock {\em Phys. Rev. E}, 58(1):738, 1998.

\bibitem{Rauzan2018}
B.M. Rauzan, A.Z. Nelson, S.E. Lehman, R.H. Ewoldt, and R.G. Nuzzo.
\newblock {Particle-free emulsions for 3D printing elastomers}.
\newblock {\em Adv. Func. Mater.}, 28(21):1707032, 2018.

\bibitem{ChaiRHE_ARFM}
R.~H. Ewoldt and C.~Saengow.
\newblock Designing complex fluids.
\newblock {\em Annu. Rev. Fluid Mech.}, 54:413--441, 2022.

\bibitem{SSRHE_DFD2019}
S.~Sen and R.~H. Ewoldt.
\newblock Drop impact of extensible yield-stress fluids.
\newblock In {\em APS Division of Fluid Dynamics Meeting Abstracts}, APS
  Meeting Abstracts, page Q24.008, November 2019.

\bibitem{keshavarz2012elastic}
B.~Keshavarz, S.~I. Green, and D.~T. Eadie.
\newblock Elastic liquid jet impaction on a high-speed moving surface.
\newblock {\em AIChE J.}, 58(11):3568--3577, 2012.

\bibitem{hsu2011role}
T.~T. Hsu, T.~W. Walker, C.~W. Frank, and G.~G. Fuller.
\newblock Role of fluid elasticity on the dynamics of rinsing flow by an
  impinging jet.
\newblock {\em Physics of Fluids}, 23(3):033101, 2011.

\bibitem{Sen_PhDThesis}
S.~Sen.
\newblock {\em Meaningful descriptions of thixotropy and extensibility for
  yield-stress fluid drop impact on thin films}.
\newblock PhD thesis, University of Illinois Urbana-Champaign, 2022.

\bibitem{AZN_SM2017}
A.~Z. Nelson and R.~H. Ewoldt.
\newblock Design of yield-stress fluids: a rheology-to structure inverse
  problem.
\newblock {\em Soft Matter}, 13:7578--7594, 2017.

\bibitem{Carbopol_Ovarlez}
G.~Ovarlez, Q.~Barral, and P.~Coussot.
\newblock Three-dimensional jamming and flows of soft glassy materials.
\newblock {\em Nat. Mater.}, 9:115--119, 2010.

\bibitem{shaukat2012shear}
A.~Shaukat, M.~Kaushal, A.~Sharma, and Y.~M. Joshi.
\newblock Shear mediated elongational flow and yielding in soft glassy
  materials.
\newblock {\em Soft Matter}, 8(39):10107--10114, 2012.

\bibitem{sica2020mises}
L.~U. Sica, P.~R. de~Souza~Mendes, and R.~L. Thompson.
\newblock Is the von mises criterion generally applicable to soft solids?
\newblock {\em Soft Matter}, 16(32):7576--7584, 2020.

\bibitem{Kamani_PRL2021}
K.~Kamani, G.~J. Donley, and S.~A. Rogers.
\newblock Unification of the rheological physics of yield stress fluids.
\newblock {\em Phys. Rev. Lett.}, 126:218002, 2021.

\bibitem{LuuForterre2009}
L.-H. Luu and Y.~Forterre.
\newblock Drop impact of yield-stress fluids.
\newblock {\em J. Fluid Mech.}, 632:301--327, 2009.

\bibitem{LuuForterrePRL2013}
L.-H. Luu and Y.~Forterre.
\newblock {Giant Drag Reduction in Complex Fluid Drops on Rough Hydrophobic
  Surfaces}.
\newblock {\em Phys. Rev. Lett.}, 110:184501, 2013.

\bibitem{BCB_PhysFluids2015}
B.~C. Blackwell, M.~E. Deetjen, J.~E. Gaudio, and R.~H. Ewoldt.
\newblock Sticking and splashing in yield-stress fluid drop impacts on coated
  surfaces.
\newblock {\em Phys. Fluids}, 27:043101, 2015.

\bibitem{SSRHE_JFM2020}
S.~Sen, A.~G. Morales, and R.~H. Ewoldt.
\newblock Viscoplastic drop impact on thin films.
\newblock {\em J. Fluid Mech.}, 891:A27, 2020.

\bibitem{Carbopol_Cates}
P.~Sollich, F.~Lequeux, P.~H\'ebraud, and M.~E. Cates.
\newblock Rheology of soft glassy materials.
\newblock {\em Phys. Rev. Lett.}, 78:2020--2023, 1997.

\bibitem{Carbopol_Kim}
J.-Y. Kim, J.-Y. Song, E.-J. Lee, and S.-K. Park.
\newblock Rheological properties and microstructures of carbopol gel network
  system.
\newblock {\em Coll. Polym. Sci.}, 281:614--623, 2003.

\bibitem{Bailey1959}
F.~E. Bailey, Jr. and R.~W. Callard.
\newblock {Some Properties of Poly(ethylene oxide) in Aqueous Solution}.
\newblock {\em J. Appl. Polym. Sci.}, 1(1):56--62, 1959.

\bibitem{Bailey_book}
F.~E. Bailey and J.~V. Koleske.
\newblock {\em Poly(ethylene oxide)}.
\newblock Academic Press New York, 1976.

\bibitem{Hossain2023}
M.~T. Hossain and R.~H. Ewoldt.
\newblock Protorheology.
\newblock {\em submitted}, 2023.

\bibitem{TanRHE}
M.~T. Hossain and R.~H. Ewoldt.
\newblock Do-it-yourself rheometry.
\newblock {\em Phys. Fluids}, 34:053105, 2022.

\bibitem{YaoMcKinley_JNNFM1998}
M.~Yao, G.~H. McKinley, and B.~Debbaut.
\newblock {Extensional deformation, stress relaxation and necking failure of
  viscoelastic filaments}.
\newblock {\em J. Non-Newtonian Fluid Mech.}, 79:469--501, 1998.

\bibitem{McKinleySridhar_AnnuRev2002}
G.~H. McKinley and T.~Sridhar.
\newblock Filament-stretching rheometry of complex fluids.
\newblock {\em Annu. Rev. Fluid Mech.}, 34:374--415, 2002.

\bibitem{koeppel2018extensional}
A.~Koeppel, P.~R. Laity, and C.~Holland.
\newblock Extensional flow behaviour and spinnability of native silk.
\newblock {\em Soft Matter}, 14(43):8838--8845, 2018.

\bibitem{McKinleyHassager_JoR1999}
G.~H. McKinley and O.~Hassager.
\newblock {The Consid\'ere condition and rapid stretching of linear and
  branched polymer melts}.
\newblock {\em J. Rheol.}, 43(5):1195--1212, 1999.

\bibitem{BachHassager_JNNFM2002}
A.~Bach, H.~K. Rasmussen, P.-Y. Longin, and O.~Hassager.
\newblock {Growth of non-axisymmetric disturbances of the free surface in the
  filament stretching rheometer: experiments and simulation}.
\newblock {\em J. Non-Newtonian Fluid Mech.}, 108:163--186, 2002.

\bibitem{BachHassager_JoR2003}
A.~Bach, H.~K. Rasmussen, and O.~Hassager.
\newblock {Extensional viscosity for polymer melts measured in the filament
  stretching rheometer}.
\newblock {\em J. Rheol}, 47(2):429--441, 2003.

\bibitem{DPL_vol1}
R.~B. Bird, R.~C. Armstrong, and O.~Hassager.
\newblock {\em Dynamics of Polymeric Liquids}, volume~1.
\newblock Wiley, 1987.

\bibitem{Macosko:1994}
C.~W. Macosko.
\newblock {\em {Rheology : principles, measurements, and applications}}.
\newblock Wiley-VCH, New York, 1994.

\bibitem{DinkgreveBonn2015}
M.~Dinkgreve, J.~Paredes, M.~A.~J. Michels, and D.~Bonn.
\newblock Universal rescaling of flow curves for yield-stress fluids close to
  jamming.
\newblock {\em Phys. Rev. E}, 92:012305, 2015.

\bibitem{Caggioni2020}
M.~Caggioni, V.~Trappe, and P.~T. Spicer.
\newblock {Variations of the Herschel–Bulkley exponent reflecting
  contributions of the viscous continuous phase to the shear rate-dependent
  stress of soft glassy materials}.
\newblock {\em J. Rheol.}, 64:413--422, 2020.

\bibitem{DonleyRogers_JNNFM2019}
G.~J. Donley, J.~R. dr~Bruyn, G.~H. McKinley, and S.~A. Rogers.
\newblock {Time-resolved dynamics of the yielding transition in soft
  materials}.
\newblock {\em J. Non-Newtonian Fluid. Mech.}, 264:117--134, 2019.

\bibitem{AsheshGaurav_SM2019}
A.~Ghosh, G.~Chaudhary, J.~G. Kang, P.~V. Braun, R.~H. Ewoldt, and K.~S.
  Schweizer.
\newblock {Linear and nonlinear rheology and structural relaxation in dense
  glassy and jammed soft repulsive pNIPAM microgel suspensions}.
\newblock {\em Soft Matter}, 15:1038--1052, 2019.

\bibitem{Valette2019}
R.~Valette, E.~Hachem, M.~Khalloufi, A.~S. Pereira, M.~R. Mackley, and S.~A.
  Butler.
\newblock {The effect of viscosity, yield stress, and surface tension on the
  deformation and breakup profiles of fluid filaments stretched at very high
  velocities}.
\newblock {\em J. Non-Newtonian Fluid Mech.}, 263:130--139, 2019.

\bibitem{Tsamopoulos_EVP2020}
P.~Moschopoulos, A.~Syrakos, Y.~Dimakopoulos, and J.~Tsamopoulos.
\newblock Dynamics of viscoplastic filament stretching.
\newblock {\em J. Non-Newtonian Fluid Mech.}, 284:104371, 2020.

\bibitem{Callister}
W.~D. Callister and D.~G. Rethwisch.
\newblock {\em Materials Science and Engineering}.
\newblock John Wiley \& Sons, 10 edition, 2020.

\bibitem{James_RheolActa2006}
D.~F. James and N.~Yogachandran.
\newblock {Filament-breaking length - a measure of elasticity in extension}.
\newblock {\em Rheol. Acta}, 45:161--170, 2006.

\bibitem{Fielding_PRL2011}
S.~M. Fielding.
\newblock {Criterion for extensional necking instability in polymeric fluids}.
\newblock {\em Phys. Rev. Lett.}, 107:258301, 2011.

\bibitem{HoyleFielding_JoR2016_1}
D.~M. Hoyle and S.~M. Fielding.
\newblock {Criteria for extensional necking instability in complex fluids and
  soft solids. Part I: Imposed Hencky strain rate protocol}.
\newblock {\em J. Rheol.}, 60(6):1347--1375, 2016.

\bibitem{HoyleFielding_JNNFM2017}
D.~M. Hoyle and S.~M. Fielding.
\newblock {Necking after extensional filament stretching of complex fluids and
  soft solids}.
\newblock {\em J. Non-Newtonian Fluid Mech.}, 247:132--145, 2017.

\bibitem{ThompsonSoares2016}
R.~L. Thompson and E.~J. Soares.
\newblock Viscoplastic dimensionless numbers.
\newblock {\em J. Non-Newtonian Fluid Mech.}, 238:57--64, 2016.

\bibitem{JalaalLohse_JFM2019}
M.~Jalaal, D.~Kemper, and D.~Lohse.
\newblock Viscoplastic water entry.
\newblock {\em J. Fluid Mech.}, 864:596--613, 2019.

\bibitem{Ashby_book2011}
M.~F. Ashby.
\newblock {\em Materials Selection in Mechanical Design}.
\newblock Elsevier, 4 edition, 2011.

\bibitem{Saramito_JNNFM2007}
P.~Saramito.
\newblock A new constitutive equation for elastoviscoplastic fluid flows.
\newblock {\em J. Non-Newtonian Fluid Mech.}, 145:1--14, 2007.

\bibitem{Saramito_JNNFM2009}
P.~Saramito.
\newblock A new elastoviscoplastic constitutive model based on the
  {Herschel-Bulkley} viscoplastic model.
\newblock {\em J. Non-Newtonian Fluid Mech.}, 158:154--161, 2009.

\bibitem{SethBonnecaze_NatMat2011}
J.R. Seth, L.~Mohan, C.~Locatelli-Champagne, M.~Cloitre, and R.T. Bonnecaze.
\newblock {A micromechanical model to predict the flow of soft particle
  glasses}.
\newblock {\em Nature Mater.}, 10:838--843, 2011.

\bibitem{RHE_JoR2013}
R.~H. Ewoldt.
\newblock Defining nonlinear rheological material functions for oscillatory
  shear.
\newblock {\em J. Rheol.}, 57:177--195, 2013.

\bibitem{RHE_baddata}
R.~H. Ewoldt, M.~T. Johnston, and L.~M. Caretta.
\newblock Experimental challenges of shear rheology: how to avoid bad data.
\newblock In S.~Spagnolie, editor, {\em Complex Fluids in Biological Systems},
  pages 207--241. Springer Biological Engineering Series, 2015.

\bibitem{Ewoldt2008}
R.~H. Ewoldt.
\newblock New measures for characterizing nonlinear viscoelasticity in large
  amplitude oscillatory shear.
\newblock {\em J. Rheol.}, 52:1427--1458, 2008.

\bibitem{Hyun2002}
K.~Hyun, S.~H. Kim, K.~H. Ahn, and S.~J. Lee.
\newblock Large amplitude oscillatory shear as a way to classify the complex
  fluids.
\newblock {\em J. Non-Newtonian Fluid Mech.}, 107:51--65, 2002.

\bibitem{Donley_PNAS2020}
G.~J. Donley, P.~K. Singh, A.~Shetty, and S.~A. Rogers.
\newblock {Elucidating the $G^{\prime\prime}$ overshoot in soft materials with
  a yield transition via a time-resolved experimental strain decomposition}.
\newblock {\em Proc. Natl. Acad. Sci. U.S.A.}, 117:21945--21952, 2020.

\bibitem{Boger-Boger}
D.~V. Boger.
\newblock A highly elastic constant-viscosity fluid.
\newblock {\em J. Non-Newtonian Fluid Mech.}, 3:87--91, 1977/1978.

\bibitem{JamesARFM2009}
D.~F. James.
\newblock Boger fluids.
\newblock {\em Annu. Rev. Fluid Mech.}, 41:129--142, 2009.

\bibitem{Moller2009}
P.~Moller, A.~Fall, V.~Chikkadi, D.~Derks, and D.~Bonn.
\newblock An attempt to categorize yield stress fluid behaviour.
\newblock {\em Philos. Trans. R. Soc. A Math. Phys. Eng. Sci.}, 367:5139--5155,
  2022.

\bibitem{Gordon2017}
M.~B. Gordon, C.~J. Kloxin, and N.~J. Wagner.
\newblock The rheology and microstructure of an aging thermoreversible
  colloidal gel.
\newblock {\em J. Rheol.}, 61:23--34, 2017.

\bibitem{Ong2019}
E.~E. Ong, S.~O'Byrne, and J.~L. Liow.
\newblock Yield stress measurement of a thixotropic colloid.
\newblock {\em Rheol. Acta}, 58:383--401, 2019.

\bibitem{SSJOR2022}
S.~Sen and R.~H. Ewoldt.
\newblock Thixotropic spectra and ashby-style charts for thixotropy.
\newblock {\em J. Rheol.}, 66(5):1041--1053, 2022.

\bibitem{Appel_PNAS2016}
A.C. Yu, H.~Chen, D.~Chan, G.~Agmon, L.M. Stapleton, A.M. Sevit, M.W. Tibbitt,
  J.D. Acosta, T.~Zhang, P.W. Franzia, R.~Langer, and E.A. Appel.
\newblock Scalable manufacturing of biomimetic moldable hydrogels for
  industrial applications.
\newblock {\em Proc. Natl. Acad. Sci. U.S.A.}, 113:14255--14260, 2016.

\bibitem{Poole_SM2019}
A.~Corker, H.~C.-H. Ng, R.~J. Poole, and E.~Garc\'ia-Tu\~n\'on.
\newblock {3D printing with 2D colloids: designing rheology protocols to
  predict `printability' of soft-materials}.
\newblock {\em Soft Matter}, 15:1444--1456, 2019.

\bibitem{disease_Zayas}
G.~Zayas, M.~C. Chiang, E.~Wong, F.~MacDonald, C.~F. Lange, A.~Senthilselvan,
  and M.~King.
\newblock Cough aerosol in healthy participants: fundamental knowledge to
  optimize droplet-spread infectious respiratory disease management.
\newblock {\em B.M.C. Pulmonary Med.}, 12:11, 2012.

\bibitem{disease_Stone}
M.~Abkarian and H.~A. Stone.
\newblock Stretching and break-up of saliva filaments during speech: A route
  for pathogen aerosolization and its potential mitigation.
\newblock {\em Phys. Rev. Fluids}, 5:102301, 2020.

\bibitem{texture_Stokes}
J.~R. Stokes, M.~W. Boehm, and S.~K. Baier.
\newblock Oral processing, texture and mouthfeel: From rheology to tribology
  and beyond.
\newblock {\em Curr. Opin. Solid State Mater. Sci.}, 18:349--359, 2013.

\bibitem{texture_Nicolas}
Y.~Nicolas and M.~Paques.
\newblock Microrheology: An experimental technique to visualize food structure
  behavior under compression-extension deformation conditions.
\newblock {\em J. Food Sci.}, 68:1990--1994, 2003.

\bibitem{texture_Fischer}
P.~Fischer and E.~J. Windhab.
\newblock Rheology of food materials.
\newblock {\em Curr. Opin. Solid State Mater. Sci.}, 16:36--40, 2011.

\end{thebibliography}

\end{document}